\documentclass[journal=jpcl,manuscript=paper,layout=twocolumn]{achemso}
\setkeys{acs}{articletitle = true}
\usepackage{color}

\usepackage{graphicx}
\usepackage{dcolumn}
\usepackage{bm}

\usepackage[utf8]{inputenc}
\usepackage[T1]{fontenc}
\usepackage{mathptmx}
\usepackage{stmaryrd}
\usepackage{relsize}
\usepackage{amsmath}

\title[An \textsf{achemso} demo]
{\large Interference Between Molecular and Photon Field-Mediated Electron Transfer Coupling Pathways in Cavities}

\author{\small Sutirtha N. Chowdhury}
\email{sutirtha.chowdhury@duke.edu}
\affiliation{\small Department of Chemistry, Duke University, Durham, North Carolina 27708, USA}

\author{\small Peng Zhang}
\email{peng.zhang@duke.edu}
\affiliation{\small Department of Chemistry, Duke University, Durham, North Carolina 27708, USA}

\author{\small David N. Beratan}
\email{david.beratan@duke.edu}
\affiliation{\small Department of Chemistry and
Department of Physics, Duke University, Durham, North Carolina 27708, USA; Department of Biochemistry, Duke University, Durham, North Carolina 27710, USA}%

\date{\today}

\begin{document}
\begin{abstract}
{\footnotesize Cavity polaritonics is capturing the imagination of the chemistry community because of the novel opportunities it creates to direct chemistry.  Electron transfer (ET) reactions are among the simplest reactions, and they also  underpin bioenergetics.  As such, new conceptual strategies to manipulate and direct electron flow at the nanoscale are of wide ranging interest in biochemistry, energy science, bio-inspired materials science, and chemistry. We show that optical cavities can modulate  electron transfer pathway interferences and ET rates in donor-bridge-acceptor (DBA) systems. We derive the  rate for DBA electron transfer systems when they are coupled with  cavity photon fields (which may be off-  or on-resonance with a molecular electronic transition), emphasizing novel cavity-induced pathway interferences with the molecular electronic  coupling pathways, as these interferences allow a new kind of ET rate tuning. We also examined the ET kinetics for both low and high cavity frequency regimes as the light-matter coupling strength is varied. The interference between the cavity-induced and intrinsic molecular coupling pathway interference is defined  by the cavity properties, including the cavity frequency and the light-matter coupling interaction strength. Thus, manipulating the cavity-induced interferences with the   chemical coupling pathways offers new strategies to direct charge flow at the nanoscale.}  
\begin{figure}
 \centering
  \begin{minipage}[t]{0.8\linewidth}
     \centering
     \includegraphics[width=\linewidth]{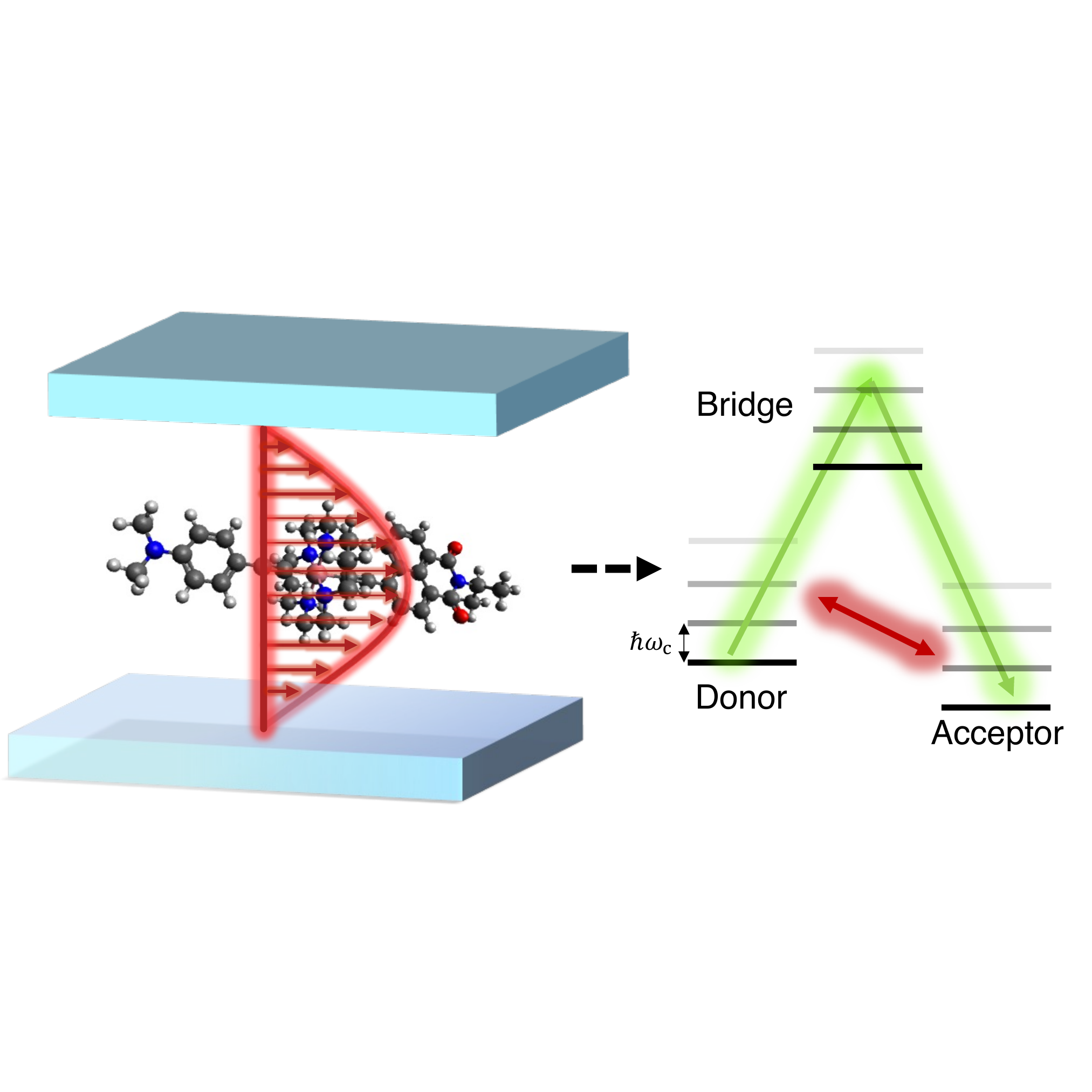}
  \end{minipage}%
\label{fig:TOC}
\end{figure}
\end{abstract}

{\small Coupling molecular system to a photon field in an optical cavity allows new strategies to redirect chemistry~\cite{Hutchison12,Thomas2019,Anoop2020,Kowalewski2016,KowalewskiJCP2016,Feist2018,Herrera16,Triana2018, Mandal2019JPCL,DU2019,Bing2020,Galego2019,Rubio2018PF,Bennett2016,Fregoni2018,Vendrell2018,csehi2019ultrafast,Szidarovszky18,Groenhof2018,Groenhof2019,Rubio2018JPB,li2022semiclassical, xinyang_natcomm_2021, xinyang_jpcl_2021, mandal_2022_JPCL,engelhardt2022unusual,lindoy2022,sun2022suppression}. Theory has played an important role in defining the principles of cavity quantum-electrodynamics (cQED) that enable new reactivities in a photon-molecule hybrid system. Despite encouraging progress, understanding electron transfer (ET) mechanisms in molecules coupled to an optical cavity is challenging. Previous theoretical studies~\cite{Herrera16,Nitzan2019,Mandal2020,chowdhury2021_qed,saller2022} addressed cavity effects on ET activation free energies and nonadiabatic interactions that arise when coupling the cavity mode to local donor excitations, or with the donor-to-acceptor transitions. These earlier studies indicate that ET rates can be altered significantly, through their effective Franck-Condon factors, when a molecule is coupled to a cavity.

Our goal is to understand how cavity interactions may modulate electron tunneling from donor to acceptor and, in particular, how the cavity may alter the coupling {\it pathway interference} in donor-bridge-acceptor (DBA) systems. In  bridge-mediated ET with off-resonant bridges, the donor-acceptor coupling arises from a combination of through-bond and through-space interactions, and ET occurs via superexchange mediated by the many coupling pathways established by these interactions. \cite{beratan_2019_ARPC,spiros_2008_prl,spiros_2013_biopolymers,gray_winkler_2003} Interestingly, we find that the cavity modulates the effective donor-acceptor (DA) interaction and the nature of the interference between direct (cavity mediated) and superexchange coupling pathways. Without light-matter coupling, we use a Marcus-like ~\cite{marcus1956}  nonadiabatic ET rate theory to describe the bridge-mediated ET. We further derive a rate expression in the weak coupling (nonadiabatic) limit for the hybrid cavity DBA system when the cavity mode is off- or on-resonance with  molecular electronic transitions. In particular, we explore thermal ET in mixed valence (MV) DBA compounds when the cavity mode is resonantly coupled to the 
intervalence charge-transfer electronic excitation (IV-CT)~\cite{Heckmann2012,dereka2016direct, londergan2002solvent,launay_2020}. The IV-CT band is an optically induced charge transfer transition that shifts the electron from D to A.~\cite{hush1968homogeneous,hopfield1977photo,richardson1984mixed} 

We find that the optical cavity creates a family of field-mediated donor-acceptor coupling pathways that interfere with the  superexchange paths that are intrinsic to the DBA structure in the absence of the cavity. Importantly, the cavity tunes the strength of the interference between the two classes of coupling pathways. By changing the properties of the cavity, including the photon frequency, light-matter coupling strength, and quantum state of the photon, we show that one can plausibly suppress or enhance the rate of ET in DBA structures. This finding suggests the possibility of modulating coupling pathways and ET rates in optical cavities, creating a new strategy for ET rate manipulation.~\cite{skourtis2010fluctuations}

We use the Pauli-Fierz (PF) non-relativistic quantum electrodynamics (QED) Hamiltonian~\cite{Cohen-Tannoudji,Rubio2018JPB,Flick2017PNAS,Vendrell2018,Mandal2020} $\hat{H}_{\mathrm{PF}}$ to describe the DBA system (electrons and nuclei) $\hat{H}_\mathrm{M}$ coupled to the radiation field $\hat{H}_\mathrm{p}=(\hat{a}^{\dagger}\hat{a}+\frac{1}{2})\hbar\omega_\mathrm{c}$ through $\hat{H}_{\mathrm{int}}$. For a molecule coupled to a single photon mode inside an optical cavity in the long wavelength limit,~\cite{Rubio2018JPB} the PF Hamiltonian is
\begin{eqnarray}\label{eqn:PF_Ham}
\hat{H}_\mathrm{PF} &=& \hat{H}_\mathrm{M} + \hat{H}_{\mathrm{P}} + \hat{H}_{\mathrm{int}}\\
&=&\hat{H}_M +  (\hat{a}^{\dagger}\hat{a}+\frac{1}{2})\hbar\omega_\mathrm{c} + \hat{\pmb{\chi}}\cdot{\hat{\pmb {\mu}}}(\hat{a}^{\dagger} + \hat{a}) + \frac{(\hat{\pmb{\chi}} \cdot{\hat{\pmb{\mu}}})^2}{\hbar\omega_\mathrm{c}}\nonumber\\
&=&\hat{H}_{\mathrm{M}}+\frac{1}{2}\hat{P}_\mathrm{c}^{2}+\frac{1}{2}\omega_\mathrm{c}^{2}\Big(\hat{Q}_\mathrm{c} + \sqrt{\frac{2}{\hbar\omega_c^3}} \hat{\pmb{ \chi}}\cdot{\hat{\pmb{\mu}}}\Big)^2\nonumber
\end{eqnarray}
$\hat{a}^{\dagger}$ and $\hat{a}$ are the  photon creation and annihilation operators respectively. $\hat{Q}_\mathrm{c} = \sqrt{\hbar/2\omega_\mathrm{c}}(\hat{a}^{\dagger} + \hat{a})$ and $\hat{P}_\mathrm{c} = i\sqrt{\hbar\omega_\mathrm{c}/2}( \hat{a}^{\dagger} - \hat{a})$ are the  photon coordinate and momentum operators, where $\omega_\mathrm{c}$ is the photon frequency in the cavity. $\hat{H}_{\mathrm{int}} = \hat{\pmb{ \chi}}\cdot{\hat{\pmb{ \mu}}}(\hat{a}^{\dagger} + \hat{a})+\frac{(\hat{\pmb {\chi}} \cdot{\hat{\pmb{\mu}}})^2}{\hbar\omega_\mathrm{c}}$ describes the light-matter interaction, where, $\hat{\pmb{\chi}} = \sqrt{\frac{\hbar \omega_\mathrm{c}}{2\epsilon_0 \mathcal{V}}}{\hat{\bf e}}$. The unit vector $\hat{{\bf e}}$ is along the polarization direction, and $\mathcal{V}$ is the quantization volume for the cavity-photon field. $\varepsilon_0$ is the permittivity in the cavity. Finally, $\pmb{\hat{\mu}}$ is the molecular dipole operator (for both electrons and nuclei).

Previous findings indicate that the coupling with an optical cavity can significantly change ET rates by influencing the Franck-Condon factor.~\cite{Herrera16,Joel_2019_natcomm,Nitzan2019,Mandal2020,chowdhury2021_qed,saller2022,xinyang_jpcl_2021,xinyang_natcomm_2021,mandal_2022_JPCL,yang_jpcl_2021,DU2021,Galego2019} However,  prior studies did not explore how the cavity might alter the bridge-mediated  (superexchange) couplings or their interferences with the direct cavity-mediated interaction between D and A.  We study a DBA system coupled to one radiation mode in an optical cavity. The DBA molecular Hamiltonian $\hat{H}_\mathrm{M}$ is:
\begin{eqnarray}\label{eqn:H_e}
&&\hat{H}_\mathrm{M} = \hat{T}_{\mathrm{s}} +\sum_{i} U_{i}|i\rangle\langle i| + V_{\mathrm{DB}}(|\mathrm{D}\rangle\langle \mathrm{B}| + |\mathrm{B}\rangle \langle \mathrm{D}|)\\
&&~+  V_{\mathrm{BA}}(|\mathrm{B}\rangle\langle \mathrm{A}| + |\mathrm{A}\rangle \langle \mathrm{B}|)+ \sum_{i} \frac{1}{2}M_{\mathrm{s}}\omega_{\mathrm{s}}^2(R_{\mathrm{s}}-R_{i}^{0})^2|i\rangle\langle i| \nonumber\\ 
&&~+\hat{H}_{\mathrm{sb}},\nonumber
\end{eqnarray}
\begin{figure}[h!]
 \centering
  \begin{minipage}[t]{1.0\linewidth}
     \centering
     \includegraphics[width=\linewidth]{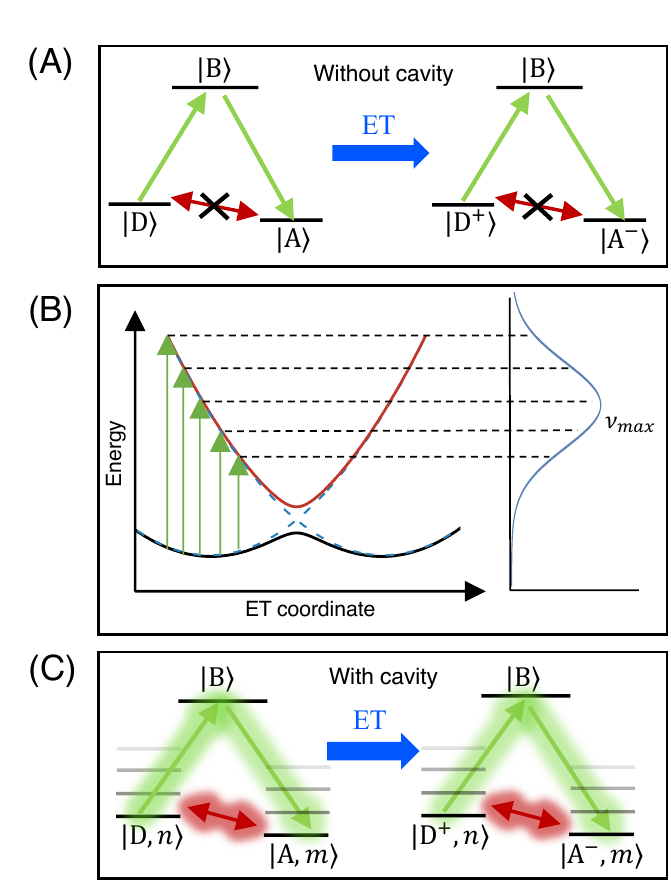}
  \end{minipage}%
   \caption{\footnotesize  (A) Schematic energy landscape for a DBA-ET structure without a cavity. Electron tunneling from $\mathrm{D}$ to $\mathrm{A}$ through the  bridge is described using an effective two-level Hamiltonian as illustrated in panel (B).  Green arrows indicate bridge-mediated interactions, and no direct interactions are present without the cavity. (B) A typical potential energy surface for the IV-CT DBA complex and its absorption spectra, which are resonantly coupled with the cavity photon field. $\nu_{max}$ is the frequency of the absorption band maxima. (C) Light-matter hybrid DBA structure, where $|\mathrm{D},n\rangle$ and $|\mathrm{A},m\rangle$ are the donor and acceptor photon dressed states, respectively. The green arrows represent the bridge-mediated interactions and the red arrow indicates the direct DA interactions in the light-matter hybrid manifold.}
\label{fig:dba_system_ivct}
\end{figure}
where $|i\rangle \in \{|\mathrm{D}\rangle, |\mathrm{B}\rangle, |\mathrm{A}\rangle\}$ indexes the diabatic donor, bridge, and acceptor states, $\hat{T}_{\mathrm{s}} = \hat{P}_{\mathrm{s}}^2/2M_{\mathrm{s}}$ is the kinetic energy operator of the reaction coordinate, $R_{\mathrm{s}}$, centered at $R_{i}^{0}$, with nuclear mass
$M_{\mathrm{s}}$ and frequency $\omega_{\mathrm{s}}$. Further, $U_{i}$ is the (constant) diabatic state energy (site energy) associated with state $|i\rangle$. $V_{\mathrm{DB}}$ and $V_{\mathrm{BA}}$ are the constant diabatic couplings of donor with bridge and bridge with acceptor, respectively. The direct diabatic coupling between $|\mathrm{D}\rangle$ and $|\mathrm{A}\rangle$ is neglected ${\it i.e.}$, $V_{\mathrm{DA}}=V_{\mathrm{AD}}=0$ as the donor to acceptor distance is typically larger than $1$ nm. The solvent reorganization energy is defined as $\lambda_{\mathrm{DA}}  = \frac{1}{2}M_{\mathrm{s}}\omega_{\mathrm{s}}^2(R_{\mathrm{D}}^0 - R_{\mathrm{A}}^0)^2$. We take $R_{\mathrm{D}}^{0} = 0$ and $R_{\mathrm{A}}^{0}=\sqrt{2\lambda_{\mathrm{DA}}/f_0}$, where $f_0$ is the force constant and $\omega_{\mathrm{s}}= \sqrt{f_0/M_{\mathrm{s}}}$. We used a solvent reorganization energy of $\lambda_{\mathrm{DA}}$ = 0.65 eV and we assume degenerate donor and acceptor states.  That is, the site-energy difference between donor and acceptor states is  $U_{\mathrm{D}}-U_{\mathrm{A}}$ = 0. Also, the site-energy difference between the (off resonance) bridge and the donor is $\Delta E = U_{\mathrm{B}}-U_{\mathrm{D}}$ = 1.5 eV.  $\hat{H}_{\mathrm{sb}}$ is the system-bath interaction Hamiltonian.~\cite{chowdhury2021_qed} The above molecular system Hamiltonian is characteristic of  type-II MV compounds, where the effective $\mathrm{DA}$ coupling is less than the reorganization energy $\lambda$.~\cite{Heckmann2012,launay_2020} 

For type-II MV structures of interest, we assume that both the transition and permanent dipole moments of the molecule are constant, ${\it i.e.}$, they are not a function of the solvent polarization coordinate. We assume that these dipoles are aligned with the cavity field polarization direction $\hat{{\bf e}}$ and that the light field couples directly with the IV-CT electronic transition. Hence, the light-matter interaction is 
\begin{eqnarray}\label{eqn:pojected_dipole_op}
&&\hat{\pmb{\chi}}\cdot\pmb{\hat{\mathrm{\mu}}} = \chi \mu_{\mathrm{DA}}(|\mathrm{D}\rangle\langle \mathrm{A}|+|\mathrm{A}\rangle\langle \mathrm{D}|) +\chi\mu_{\mathrm{DD}}|\mathrm{D}\rangle\langle \mathrm{D}|\nonumber\\
&&~+ \chi\mu_{\mathrm{AA}}|\mathrm{A}\rangle\langle \mathrm{A}|
\end{eqnarray}
where, $\chi = \sqrt{\frac{\hbar \omega_{\mathrm{c}}}{2\varepsilon_0 \mathcal{V}}}$ and  we further assume that both $\hat{\pmb{\chi}}$ and $\hat{\pmb{\mu}}$ are oriented along the cavity field polarization direction $\hat{\mathbf{e}}$. Using Eq.~\ref{eqn:pojected_dipole_op}, the light-matter interaction Hamiltonian $\hat{H}_{\mathrm{int}}$, is  
\begin{eqnarray}\label{eqn:H_int}
\hat{H}_{\mathrm{int}} &=& \hbar g_\mathrm{c}(|\mathrm{D}\rangle\langle \mathrm{A}|+|\mathrm{A}\rangle\langle \mathrm{D}|)(\hat{a}^{\dagger}+\hat{a}) + (\chi \mu_{\mathrm{DD}}|\mathrm{D}\rangle\langle \mathrm{D}|\nonumber\\
&&~+ \chi  \mu_{\mathrm{AA}}|\mathrm{A}\rangle\langle \mathrm{A}|)(\hat{a}^{\dagger} + \hat{a}) +\frac{(\hat{\pmb{\chi}}\cdot\hat{\pmb{\mu}})^2}{\hbar \omega_{\mathrm{c}}},
\end{eqnarray}
where the coupling strength $\hbar g_\mathrm{c}\equiv \chi{{\mu}}_{\mathrm{DA}}$, and  $\frac{(\hat{\pmb{\chi}}\cdot\hat{\pmb{\mu}})^2}{\hbar \omega_c}$ is the dipole-self energy (DSE). Note that,  $\hat{\pmb{\chi}}\cdot \hat{\pmb{\mu}}$ has units of energy, and the unit of $\hat{\pmb {\chi}}$ is energy/dipole moment. We set the transition and permanent dipole moment as unitless parameters (as we scaled by units of dipole moment in the denominator). In this model, we used the $\mathrm{D}$ to $\mathrm{A}$ transition dipole moment $\mu_{\mathrm{DA}}$ = 1 and the permanent dipole moments  $\mu_{\mathrm{DD}}=$ 5 and $\mu_{\mathrm{AA}}=$ -5.

Figure~\ref{fig:dba_system_ivct} shows a typical DBA electron-transfer system with and without the cavity, and the associated IV-CT excitation. Without the cavity, ET is mediated by superexchange via one bridge state and by the direct through-space donor-acceptor electronic coupling interaction (Figure~\ref{fig:dba_system_ivct}(A)). When the molecule is coupled to the cavity  (Figure~\ref{fig:dba_system_ivct}(C)), ET occurs via cavity-created light-matter hybrid states, such as $|\mathrm{D},n\rangle$ (the donor state with $n$ photons in the cavity), and $|\mathrm{A},m\rangle$ (the acceptor state with $m$ photons in the cavity). Each dressed channel is separated by the energy of the photon field $\hbar \omega_{\mathrm{c}}$ in the cavity. The bridge state is not dressed by the photon numbers, because the light field is only  resonant with the $|\mathrm{D}\rangle \to |\mathrm{A}\rangle$ IV-CT transition. In this cavity-enabled mechanism, ET occurs by (1)  bridge-mediated superexchange ($|\mathrm{D},n\rangle \to |\mathrm{B}\rangle \to |\mathrm{A},m\rangle$), and (2) direct through-space donor to acceptor tunneling ($|\mathrm{D},n\rangle \to |\mathrm{A},m\rangle$) mediated by cavity modes. These two classes of pathways provide very different rate profiles with respect to the light-matter interaction strength. Importantly, The ET rate depends on the interference between the superexchange and direct coupling pathways.     

We use two analytical rate theories to quantify the ET rate constants for our dressed molecular system. Without the cavity, the ET rate is described by a non-adiabatic Marcus-like rate  ~\cite{marcus1956,marcus1985}. With the cavity, the Fermi golden rule (FGR) description of the polariton mediated electron transfer (PMET) rate remains valid.~\cite{Herrera16,Nitzan2019,Joel_2019_natcomm,chowdhury2021_qed,saller2022,Mandal2020}  

{\bf (1) Marcus-like nonadiabatic ET:} The rate for the DBA system, without the cavity is:
\begin{equation}\label{eqn:normal_marcus}
k_\mathrm{MT} = \frac{2\pi|V_{\mathrm{DA}}^{\mathrm{eff}}|^2}{\hbar}\sqrt{\frac{1}{4\pi\lambda_{\mathrm{DA}}k_B T}}\mathrm{exp}\bigg[-\frac{(\Delta G+\lambda_{\mathrm{DA}})^2}{4\lambda_{\mathrm{DA}}k_BT}\bigg],    
\end{equation}
where, $V_{\mathrm{DA}}^{\mathrm{eff}}$ is the effective  coupling between $\mathrm{D}$ and $\mathrm{A}$ diabatic states.
\begin{equation}
V_{\mathrm{DA}}^{\mathrm{eff}} = -\frac{V_{\mathrm{DB}}V_{\mathrm{BA}}}{2}[\frac{1}{U_{\mathrm{B}}-U_{\mathrm{A}}}+\frac{1}{U_{\mathrm{B}}-U_{\mathrm{D}}}].   
\end{equation}
$\Delta G = (U_{\mathrm{A}}-U_{\mathrm{D}})-\frac{V_{\mathrm{BA}}^2}{(U_{\mathrm{B}}-U_{\mathrm{A}})} + \frac{V_{\mathrm{DB}}^2}{(U_{\mathrm{B}}-U_{\mathrm{D}})} $ is the effective ET driving force, $\lambda_{\mathrm{DA}}$ is the reorganization energy, where, $k_{B}$ is  Boltzmann's constant and $T$ is the temperature ($T=300K$).

{\bf (2) Fermi's golden rule for PMET:}
We now describe the  PMET rate when the photon frequency is resonant with the IV-CT band. PMET occurs from a set of photon-dressed donor states $|\mathrm{D},n\rangle$ to a set of photon dressed acceptor states $|\mathrm{A},m\rangle$ via a virtual bridge state $|\mathrm{B}\rangle$, or via direct interaction between $|\mathrm{D},n\rangle$ to $|\mathrm{A},m\rangle$ states. To  calculate the rates associated with each photon dressed channel for the two  reaction pathways, we follow earlier approaches~\cite{Herrera16,Nitzan2019,Mandal2020,chowdhury2021_qed} and use  Fermi's golden rule for the ET rates: ~\cite{nitzan1972,ulstrup1975,efrima1974}
\begin{eqnarray}\label{eqn:rate_each_channel}
k_\mathrm{FGR}&=& \sum_{n}{P}_n \sum_{m}\frac{|F_{nm}|^2}{\hbar}
\sqrt{\frac{\pi}{\lambda_{\mathrm{DA}}k_BT}}\nonumber\\
&&~\times \mathrm{exp}\bigg[-\frac{(\Delta G_{nm} + \lambda_{\mathrm{DA}} )^2}{4\lambda_{\mathrm{DA}}k_BT}\bigg],
\end{eqnarray}
where $F_{nm}$ is the effective $\mathrm{DA}$ coupling among photon dressed states 
\begin{eqnarray}\label{eqn:eff_DA_coup}
&&F_{nm} = \underbrace{\hbar g_{c}[\sqrt{n}S_{n-1,m}^{\mathrm{DA}} + \sqrt{n+1}S_{n+1,m}^{\mathrm{DA}}]}_{\text{direct}\rm\ DA\rm\ coupling}\\
&&-\underbrace{\frac{\tilde{V}_{n,0}^{\mathrm{DB}}\tilde{V}_{0,m}^{\mathrm{BA}}}{2}\bigg[\frac{1}{(U_{\mathrm{B}}-U_{\mathrm{A}})-m\hbar\omega_{\mathrm{c}}} + \frac{1}{(U_{\mathrm{B}}-U_{\mathrm{D}})-n\hbar\omega_{\mathrm{c}}}\bigg]}_{\text{bridge}\rm\ \text{mediated}\rm\ \text{coupling}}\nonumber
\end{eqnarray}
with, $\tilde{V}_{n,0}^{\mathrm{DB}}=V_{\mathrm{DB}}S_{n,0}^{\mathrm{DB}}$, $\tilde{V}_{0,m}^{\mathrm{BA}} = V_{\mathrm{BA}}S_{0,m}^{\mathrm{BA}}$ and  $S_{nm}^{\mathrm{DA}} = \langle n|e^{-i/\hbar \sqrt{2/\hbar \omega_{\mathrm{c}}^3}\chi(\mu_{\mathrm{DD}}-\mu_{\mathrm{AA}})\hat{P}_{\mathrm{c}}}|m\rangle$,
$S_{n,0}^{\mathrm{DB}} = \langle n|e^{-i/\hbar \sqrt{2/\hbar \omega_{\mathrm{c}}^3}\chi\mu_{\mathrm{DD}}\hat{P}_{\mathrm{c}}}|0\rangle$, and $S_{0,m}^{\mathrm{BA}} = \langle 0|e^{i/\hbar \sqrt{2/\hbar \omega_{\mathrm{c}}^3}\chi\mu_{\mathrm{AA}}\hat{P}_{\mathrm{c}}}|m\rangle$. The derivations of Eq.~\ref{eqn:rate_each_channel} and Eq.~\ref{eqn:eff_DA_coup}, and  the expressions for $\Delta G_{nm}$ (the effective driving force between photon-dressed donor and acceptor states), are provided in the SI. The thermal population of the corresponding cavity mode is
\begin{equation}
P_n= \frac{\mathrm{exp}[-\beta n \hbar \omega_c]}{\sum_n \mathrm{exp}[-\beta n \hbar \omega_c]},    
\end{equation}
where, $\beta = 1/k_{\mathrm{B}}T$. 
We also derive the rate for the off-resonance case (where the cavity mode is not  resonant with any electronic transitions). The effective $\mathrm{DA}$ couplings and the rate constants for the off-resonance theory appear in the SI.

We first describe the PMET (polariton mediated electron transfer) rate obtained from the FGR rate expression (Eq.~\ref{eqn:rate_each_channel}) for on-resonance conditions. Figure~\ref{fig:k_vs_gc_ivct} shows the computed PMET rates as a function of effective light-matter coupling ($g_{\mathrm{c}}/\omega_{\mathrm{c}}$) strengths with $\hbar\omega_{\mathrm{c}}$ = 860 meV (solid lines) and $\hbar \omega_{\mathrm{c}}$ = 430 meV (dashed lines) at $\Delta E$ = 1.5 eV. The colored lines represent $V_{\mathrm{DB}}$ and $V_{\mathrm{BA}}$ couplings: $V_{\mathrm{DB}}=V_{\mathrm{BA}}=$ 0.02  eV (black), $V_{\mathrm{DB}}=V_{\mathrm{BA}}=$ 0.04  eV (green), and $V_{\mathrm{DB}}=V_{\mathrm{BA}}=$ 0.06  eV (orange). 
When the molecule is decoupled from the cavity, the ET occurs via  $|\mathrm{D}\rangle \to |\mathrm{B}\rangle \to |\mathrm{A}\rangle$ superexchange and the rate is analyzed using Eq.~\ref{eqn:normal_marcus}.   

\begin{figure}[h!]
 \centering
  \begin{minipage}[t]{0.9\linewidth}
     \centering
     \includegraphics[width=\linewidth]{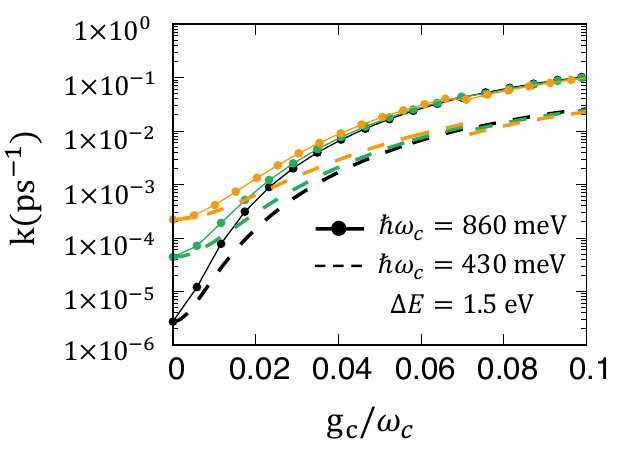}
  \end{minipage}%
   \caption{\footnotesize PMET rates constants computed for a range of effective light-matter coupling strengths ($g_{\mathrm{c}}/\omega_{\mathrm{c}}$). The solid lines correspond to  $\hbar \omega_{\mathrm{c}}$= 860 meV and the dashed lines to $\hbar\omega_{\mathrm{c}}$= 430 meV. The  colors represent different $V_{\mathrm{DB}}$ and $V_{\mathrm{BA}}$ values: $V_{\mathrm{DB}}=V_{\mathrm{BA}}=$ 0.02  eV (black), $V_{\mathrm{DB}}=V_{\mathrm{BA}}=$ 0.04  eV (green), and $V_{\mathrm{DB}}=V_{\mathrm{BA}}=$ 0.06  eV (orange). The smaller the $V_{\mathrm{DB}}$ and $V_{\mathrm{BA}}$ values, the larger the PMET rate enhancement is found to be.}
\label{fig:k_vs_gc_ivct}
\end{figure}
In the presence of a cavity, ET is mediated by two classes of coupling pathways
\begin{itemize}
    \item {\bf Pathway 1:} ET from photon dressed donor states $|\mathrm{D},n\rangle$ to photon dressed acceptor states $|\mathrm{A},m\rangle$ via a single bridge state $|\mathrm{B}\rangle$ (the bridge is not coupled with the light-field).
    \item {\bf Pathway 2:} ET from direct transitions between $|\mathrm{D},n\rangle$ and $|\mathrm{A},m\rangle$. This direct transition is determined by  cavity properties, such as the light-matter coupling strengths and the photon frequency.  
\end{itemize}
First, one finds that increasing $g_{\mathrm{c}}/\omega_{\mathrm{c}}$ causes the reaction rate to grow compared to the cavity free situation (when $g_{\mathrm{c}}/\omega_{\mathrm{c}}$ = 0). This is found because  increasing the light-matter coupling strength causes the ET to occur mostly via direct transitions between $|\mathrm{D},n\rangle$ and $|\mathrm{A},m\rangle$ (pathway 2);  the direct coupling is proportional to the light-matter coupling strength, which is  understood through  Eq.~\ref{eqn:eff_DA_coup}. The first term in Eq.~\ref{eqn:eff_DA_coup} grows with increasing light-matter coupling strength.

Second, the rate enhancement described above decreases as $V_{\mathrm{DB}}$ and $V_{\mathrm{BA}}$ couplings grow (see the black dashed/solid line and orange dashed/solid line). This finding is understood by considering the destructive interference (see Eq.~\ref{eqn:eff_DA_coup}) that arises between the bridge-mediated superexchange coupling (pathway 1), and the  direct (through-space) $\mathrm{DA}$ coupling  (pathway 2). Increasing $V_{\mathrm{DB}}$ and $V_{\mathrm{BA}}$ causes the bridge-mediated pathway to  interfere strongly with the direct (through-space) coupling from $\mathrm{D}$ to $\mathrm{A}$, so the rate enhancement is reduced compared with the case of smaller $V_{\mathrm{DB}}$ and $V_{\mathrm{BA}}$ values.


Third, in this  model, the PMET rate is larger for high cavity frequencies. The rate is nearly one order of magnitude larger when $\hbar \omega_{\mathrm{c}}$ = 860 meV compared to $\hbar \omega_{\mathrm{c}}$ = 430 meV. This effect can also be  understood as arising from coupling pathway interferences. The smaller the photon frequency, the higher energy donor dressed states ($|\mathrm{D},n\rangle$) are thermally populated, and there will be larger interactions  of donor-dressed states with the bridge state,  producing stronger destructive pathway interferences and slower ET rates. Interestingly, the cavity frequency modulated PMET rate enhancement also depends on the ET driving force (-$\Delta G_{nm}$). Previous studies~\cite{Mandal2020,chowdhury2021_qed} found that the PMET rate is significantly enhanced by a low cavity frequency in the Marcus normal regime. One observes this effect by computing the PMET rate as a function of the molecular driving force, and this finding is also consistent with the DBA model system analysis.

PMET rates may be measured using methods reported earlier,~\cite{londergan2002solvent} for example through the IR spectral broadening of  Ru(II/III) mixed-valence complexes.
Further, it is well known that ET couplings and reorganization energies can be derived from  intervalence spectra in the weak DA coupling limit. ~\cite{londergan2002solvent,Heckmann2012,hush1968homogeneous,hopfield1977photo,richardson1984mixed}  
Analogous measurement of intervalence spectra can provide information on how the donor-acceptor  couplings are influenced when the cavity mode is resonantly coupled with the IV-CT absorption band.

In addition to the resonant treatment describe above, we also developed an off-resonance rate theory for PMET (see SI). In this regime, the $\mathrm{D}$, $\mathrm{B}$, and $\mathrm{A}$ states couple with the photon field. Figure ~\ref{fig:off_resonance_1} shows the computed PMET rate found for the off-resonance regime. In this case, ET occurs from photon dressed donor states $|\mathrm{D},n\rangle$ to photon dressed acceptor states $|\mathrm{A},m\rangle$ via bridge-mediated channels $|\mathrm{B},l\rangle$ ($|\mathrm{D},n\rangle \to |\mathrm{B},l\rangle \to |\mathrm{A},m\rangle$) and also direct transition from $|\mathrm{D},n\rangle \to |\mathrm{A},m\rangle$ channels (see Figure ~\ref{fig:off_resonance_1}(A)). The key difference, relative to the resonant theory, is the contribution of the many-fold virtual bridge states that produce  stronger destructive interferences.  
\begin{figure}[h!]
 \centering
  \begin{minipage}[t]{1.0\linewidth}
     \centering
     \includegraphics[width=\linewidth]{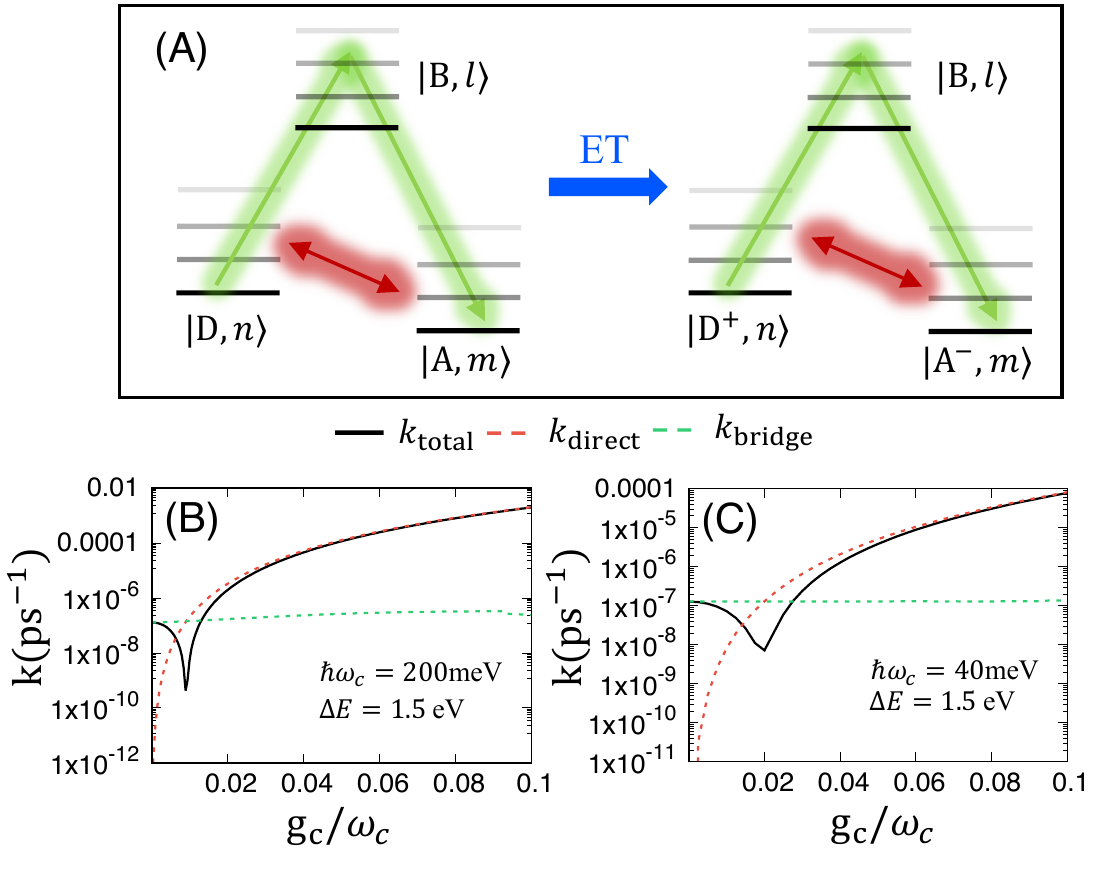}
  \end{minipage}%
   \caption{\footnotesize (A) $\mathrm{DBA}$ cavity-hybrid model system, where the cavity mode is off of resonance with the DBA electronic transition. $|\mathrm{D},n\rangle$, $|\mathrm{B},l\rangle$, and $|\mathrm{A},m\rangle$ are the photon dressed donor, bridge and acceptor states, respectively. (B) PMET rate constants calculated over a range of effective light-matter coupling strengths  ($g_{\mathrm{c}}/\omega_{\mathrm{c}}$) with  $\hbar \omega_c$ = 200 meV and (C) $\hbar \omega_c$ = 40 meV. The solid black line corresponds to the total  rate, the orange dashed line refers only to the cavity established direct (through-space) donor-to-acceptor rate, and the dashed green line represents the bridge assisted (superexchange) rate.}
\label{fig:off_resonance_1}
\end{figure}
Figure~\ref{fig:off_resonance_1}(A) shows the light-matter hybrid $\mathrm{DBA}$-ET energy landscape for  off-resonance superexchange. Figure~\ref{fig:off_resonance_1}(B) and ~\ref{fig:off_resonance_1}(C) show the computed reaction rate as a function of the effective light-matter coupling ($g_{\mathrm{c}}/\omega_{\mathrm{c}}$) at $\Delta E$ = 1.5 eV for $\hbar\omega_{\mathrm{c}}$ = 200 meV and $\hbar\omega_{\mathrm{c}}$ = 40 meV. Details of the model Hamiltonian and additional parameters appear in the SI. Each panel shows three different rate signatures, (a) the ET rate from $\mathrm{D}$ to $\mathrm{A}$, $k_{\mathrm{total}}$, (b) the ET rate for just the direct (through-space) D-A transition, $k_{\mathrm{direct}}$, (calculated  using only the direct coupling term (see Eq.~S19 in the SI)), and (c) the ET rate for just bridge mediated (superexchange) coupling, $k_{\mathrm{bridge}}$, (calculated  using only the bridge mediated coupling  (see Eq.~S20 of the SI)). In Figures~\ref{fig:off_resonance_1}(B) and ~\ref{fig:off_resonance_1}(C), the initial rate suppression is followed by a rate enhancement. This feature is explained by considering  interference between the two coupling paths. Weaker light-matter couplings cause the direct  and the bridge-mediated paths to  interfere strongly, producing rate suppression. Further  increasing the effective light-matter coupling ($g_{\mathrm{c}}/\omega_{\mathrm{c}}$), $k_{\mathrm{direct}}$ dominates over $k_{\mathrm{bridge}}$ and the overall ET rate is enhanced. The rate suppression is broader for Figure~\ref{fig:off_resonance_1}(C) compared to Figure~\ref{fig:off_resonance_1}(B). This behavior arises since, at the lower photon frequency of Figure~\ref{fig:off_resonance_1}(C), the high lying donor states are thermally populated. These states thus participate in the ET process, and they produce stronger interactions with bridge states. Hence, the enhancement of $k_{\mathrm{direct}}$  is weaker for Figure ~\ref{fig:off_resonance_1}(C), leading to a broader suppression as a function of $g_{\mathrm{c}}/\omega_{\mathrm{c}}$, and the $k_{\mathrm{direct}}$ value is almost an order of magnitude smaller,   explaining the faster growth of $k_{\mathrm{direct}}$ in Figure~\ref{fig:off_resonance_1}(B) compared to ~\ref{fig:off_resonance_1}(C). This produces two different light-matter coupling modulated ET rates in Fig.~\ref{fig:off_resonance_1}(B) and ~\ref{fig:off_resonance_1}(C).

We developed PMET rate theories for  $\mathrm{DBA}$ systems in the limit of on-resonance coupling (where the photon mode is resonant with the IV-CT band) and   off-resonance coupling. The  computed PMET rates indicate that rate modulation occurs via two coupling pathways: (a) through space paths  (cavity created, direct $|\mathrm{D}\rangle \to |\mathrm{A}\rangle$ transition) and (b) through-bond paths (cavity modulated $|\mathrm{D}\rangle \to |\mathrm{B}\rangle \to |\mathrm{A}\rangle$ transitions). We found that  ET rates can be suppressed or enhanced as a function of the effective light-matter coupling strength. This rate control is largely determined by the cavity modulated interference between the through-space and through-bond coupling pathways. This perspective on cavity-modulated coupling pathway interferences points to a  novel approach for  modulating charge flow though molecules in cavities, supplementing earlier strategies.~\cite{Mandal2020,chowdhury2021_qed,Herrera16,Nitzan2019} 

This study reveals the quantum pathway interference analysis for  $\mathrm{DBA}$ structures in a cavity. PMET is predicted to enable the tuning of ET kinetics by manipulating the interference between through-bond and through-space coupling pathways in cavities, and this approach is anticipated to provide novel strategies to manipulate the polariton chemistry of ET systems. 

Having a strategy to manipulate coupling pathway interference may also leads to new approaches to control chemical reactivity by coupling the vibrational degrees of freedom with the  radiation field, specifically in the strong vibrational coupling and nonadiabatic limits.

\section{\normalsize Acknowledgments}
{\footnotesize The authors thank the National Science Foundation (CHE-1954853) for support of this research.}

\section{\normalsize Supporting Information Available}
{\footnotesize The Supporting Information is available free of charge on the ACS Publications website.

Model system and parameters for the off resonance Hamiltonian, detailed derivation of the off-resonance rate theory, detailed derivation of the on-resonance rate theory,and additional numerical results.}

\bibliography{reference.bib}
}
\end{document}


\maketitle

\section{\normalsize Model Hamiltonian for the Off-Resonance Rate Theory}
In this section, we describe the model Hamiltonian used to describe a light-matter hybrid DBA system in off resonance conditions. As with the on resonance condition (see main text), we also begin with the Pauli-Fierz (PF) Hamiltonian. 
\begin{eqnarray}\label{eqn:PF_Ham}
\hat{H}_\mathrm{PF} &=& \hat{H}_\mathrm{M} + \hat{H}_{\mathrm{P}} + \hat{H}_{\mathrm{int}}\\
&=&\hat{H}_M +  (\hat{a}^{\dagger}\hat{a}+\frac{1}{2})\hbar\omega_\mathrm{c} + \hat{\boldsymbol \chi}\cdot{\hat{\boldsymbol \mu}}(\hat{a}^{\dagger} + \hat{a}) + \frac{(\hat{\boldsymbol \chi} \cdot{\hat{\boldsymbol \mu}})^2}{\hbar\omega_\mathrm{c}},\nonumber
\end{eqnarray}
where $\hat{a}$ and $\hat{a}^{\dagger}$ are the photon creation and annihilation operator. $\omega_{c}$ is the photon frequency in the cavity and $\hat{H}_{\mathrm{int}}$ is the light-matter interaction Hamiltonian:  
\begin{equation}
\hat{H}_{\mathrm{int}} =  \hat{\boldsymbol \chi}\cdot{\hat{\boldsymbol \mu}}(\hat{a}^{\dagger} + \hat{a}) + \frac{(\hat{\boldsymbol \chi} \cdot{\hat{\boldsymbol\mu}})^2}{\hbar\omega_\mathrm{c}}   
\end{equation}
Under off resonance conditions, the donor ($|\mathrm{D}\rangle$), bridge ($|\mathrm{B}\rangle$), and acceptor ($|\mathrm{A}\rangle$) states are all coupled with the light field and the light-matter interaction term is written:
\begin{eqnarray}\label{eqn:chi_dot_mu}
\pmb{\hat{\chi}}\cdot\pmb{\hat{\mathrm{\mu}}} &=& \chi \mu_{\mathrm{DB}}(|\mathrm{D}\rangle\langle \mathrm{B}|+|\mathrm{B}\rangle\langle \mathrm{D}|) + \chi \mu_{\mathrm{BA}}(|\mathrm{B}\rangle\langle \mathrm{A}|+|\mathrm{A}\rangle\langle \mathrm{B}|) + \chi \mu_{\mathrm{DD}}|\mathrm{D}\rangle\langle \mathrm{D}|\nonumber\\
&&~+\chi \mu_{\mathrm{BB}}|\mathrm{B}\rangle\langle \mathrm{B}| + \chi \mu_{\mathrm{AA}}|\mathrm{A}\rangle\langle \mathrm{A}|
\end{eqnarray}
where, $\chi = \sqrt{\frac{\hbar \omega_{\mathrm{c}}}{2\varepsilon_0 \mathcal{V}}}$. We set $\mu_{\mathrm{DA}}=\mu_{\mathrm{AD}}$ = 0, as charge transfer band absorption is typically weak. Using the Eq.~\ref{eqn:chi_dot_mu} the $\hat{H}_{\mathrm{int}}$ term is:
\begin{eqnarray}\label{eqn:H_int_off}
\hat{H}_{\mathrm{int}} &=& \hbar g_\mathrm{c}(|\mathrm{D}\rangle\langle \mathrm{B}|+|\mathrm{B}\rangle\langle \mathrm{D}|)(\hat{a}^{\dagger}+\hat{a})+\hbar \eta_\mathrm{c}(|\mathrm{B}\rangle\langle \mathrm{A}|+|\mathrm{A}\rangle\langle \mathrm{B}|)(\hat{a}^{\dagger}+\hat{a})\nonumber\\
&+&(\chi \mu_{\mathrm{DD}}|\mathrm{D}\rangle\langle \mathrm{D}|+ \chi \mu_{\mathrm{BB}}|\mathrm{B}\rangle\langle \mathrm{B}|+\chi \mu_{\mathrm{AA}}|\mathrm{A}\rangle\langle \mathrm{A}|)(\hat{a}^{\dagger} + \hat{a}) +\frac{(\hat{\pmb{\chi}}\cdot\hat{\pmb{\mu}})^2}{\hbar \omega_{\mathrm{c}}},
\end{eqnarray}
where the coupling strength $\hbar g_\mathrm{c}\equiv \chi{{\mu}}_{\mathrm{DB}}$, $\hbar \eta_\mathrm{c}\equiv \chi {{\mu}}_{\mathrm{BA}}$. In this light-matter hybrid system, we took  $\mu_{\mathrm{DB}}=\mu_{\mathrm{BA}} = 1$, and the permanent dipole moment differences are  $\mu_{\mathrm{DD}}-\mu_{\mathrm{AA}} =$ 1, $\mu_{\mathrm{DD}}-\mu_{\mathrm{BB}}=$ 5, and $\mu_{\mathrm{BB}}-\mu_{\mathrm{AA}}=$ 5. We treat the transition and permanent dipole moments as unitless numbers, but $\hat{\pmb{\chi}} \cdot \hat{\pmb{\mu}}$ has units of energy, and the $\mathrm{DBA}$ molecular Hamiltonian $\hat{H}_M$ is the same as in Eq.~(2) of the main text. The model parameters for $\hat{H}_M$ are exactly the same as in the on-resonance case (see main text), except that we set $V_{\mathrm{DB}} = V_{\mathrm{BA}} = 5$ meV, and the site energy difference between donor and acceptor states is $U_{\mathrm{D}}-U_{\mathrm{A}} = 150$ meV.

\section{\normalsize Derivation of the PMET Rate Theory for the Off-Resonance Case}
We use the following polaron transformation operator~\cite{Nitzan2019,chowdhury2021_qed} of the photonic degrees of freedom (DOF)
\begin{equation}
\hat{U}_\mathrm{pol}= \mathrm{exp}\Bigg[{\frac{i}{\hbar}\sqrt{2\over \hbar\omega_\mathrm{c}^3}{\chi} \sum_{j\in{\mathrm{D,B,A}}}\mu_{jj}|j \rangle \langle j|\hat{P}_\mathrm{c}}\Bigg].  
\end{equation}
The polaron transformation operator shifts the photonic coordinate,  $e^{-\frac{i}{\hbar}Q_\mathrm{0}\hat{P}_\mathrm{c}} \hat{O}(\hat{Q}_\mathrm{c}) e^{\frac{i}{\hbar}Q_\mathrm{0}\hat{P}_\mathrm{c}} = \hat{O}(\hat{Q}_\mathrm{c} - Q_\mathrm{0})$. $\hat{U}_\mathrm{pol}$ transforms  $|\mathrm{D}\rangle\langle \mathrm{B}|$ as follows:
\begin{equation}\label{sigmad-pol}
\hat{U}_\mathrm{pol}^{\dagger} |\mathrm{D}\rangle\langle \mathrm{B}|\hat{U}_\mathrm{pol} = e^{-\frac{i}{\hbar}\sqrt{\frac{2}{\hbar\omega_{\mathrm{c}}^3}} { \chi}(\mu_{\mathrm{DD}}-\mu_{\mathrm{BB}}) \hat{P}_\mathrm{c}}|\mathrm{D}\rangle\langle \mathrm{B|}\nonumber.
\end{equation}
This can be shown (with $\hat{\Pi}_\mathrm{c} = \frac{1}{\hbar}\sqrt{2\over \hbar\omega_\mathrm{c}^3}{\chi}\hat{P}_\mathrm{c}$) as follows:
\begin{align}\label{eqn:polaron_trans}
\hat{U}_\mathrm{pol}^{\dagger}|\mathrm{D}\rangle\langle\mathrm{B}|\hat{U}_\mathrm{pol} &= e^{-i\hat{\Pi}_\mathrm{c}\big(\mu_{\mathrm{DD}}|\mathrm{D}\rangle\langle \mathrm{D}|+\mu_{\mathrm{BB}}|\mathrm{B}\rangle\langle \mathrm{B}|+\mu_{\mathrm{AA}}|\mathrm{A}\rangle\langle \mathrm{A}|\big)} |\mathrm{D}\rangle\langle\mathrm{B}|e^{i\hat{\Pi}_\mathrm{c}\big(\mu_{\mathrm{DD}}|\mathrm{D}\rangle\langle \mathrm{D}|+\mu_{\mathrm{BB}}|\mathrm{B}\rangle\langle \mathrm{B}|+\mu_{\mathrm{AA}}|\mathrm{A}\rangle\langle \mathrm{A}|\big)}\nonumber\\
&= e^{{-i\hat{\Pi}_\mathrm{c}  \mu_{\mathrm{DD}}|\mathrm{D}\rangle\langle \mathrm{D}|}} e^{-{i\hat{\Pi}_\mathrm{c}\mu_{\mathrm{BB}} |\mathrm{B}\rangle\langle \mathrm{B}|}}e^{-{i\hat{\Pi}_\mathrm{c}\cdot\mu_{\mathrm{AA}} |\mathrm{A}\rangle\langle \mathrm{A}|}}|\mathrm{D}\rangle\langle \mathrm{B}|  e^{{i\hat{\Pi}_\mathrm{c} \mu_{\mathrm{DD}}|\mathrm{D}\rangle\langle \mathrm{D}|}} e^{{i\hat{\Pi}_\mathrm{c} \mu_{\mathrm{BB}} |\mathrm{B}\rangle\langle \mathrm{B}|}}e^{{i\hat{\Pi}_\mathrm{c} \mu_{\mathrm{AA}} |\mathrm{A}\rangle\langle \mathrm{A}|}} \nonumber\\
&= e^{{-i\hat{\Pi}_\mathrm{c} \mu_{\mathrm{DD}}  |\mathrm{D}\rangle\langle \mathrm{D}|}}e^{{-i\hat{\Pi}_\mathrm{c} \mu_{\mathrm{BB}} |\mathrm{B}\rangle\langle \mathrm{B}|}} \Big[\mathds{\hat{1}}_\mathrm{e} -  {i\hat{\Pi}_\mathrm{c} \mu_{\mathrm{AA}}} |\mathrm{A}\rangle\langle \mathrm{A}| + ... \Big]|\mathrm{D}\rangle\langle \mathrm{B}|  \Big[\mathds{\hat{1}}_\mathrm{e} +  {i\hat{\Pi}_\mathrm{c} \mu_{\mathrm{DD}}} |\mathrm{D}\rangle\langle \mathrm{D}| + ... \Big]\nonumber\\
&~~~\times e^{{i\hat{\Pi}_\mathrm{c} {\mu_{\mathrm{BB}}} |\mathrm{B}\rangle\langle \mathrm{B}|}}e^{{i\hat{\Pi}_\mathrm{c}\cdot {\mu_{\mathrm{AA}}} |\mathrm{A}\rangle\langle \mathrm{A}|}}\nonumber\\
&= e^{{-i\hat{\Pi}_\mathrm{c} \mu_{\mathrm{DD}}  |\mathrm{D}\rangle\langle \mathrm{D}|}}e^{{-i\hat{\Pi}_\mathrm{c} \mu_{\mathrm{BB}} |\mathrm{B}\rangle\langle \mathrm{B}|}} |\mathrm{D}\rangle\langle\mathrm{B}| e^{{i\hat{\Pi}_\mathrm{c} {\mu_{\mathrm{BB}}} |\mathrm{B}\rangle\langle \mathrm{B}|}}e^{{i\hat{\Pi}_\mathrm{c} {\mu_{\mathrm{AA}}} |\mathrm{A}\rangle\langle \mathrm{A}|}}\nonumber\\
&= e^{-i\hat{\Pi}_\mathrm{c} \mu_{\mathrm{DD}}|\mathrm{D}\rangle\langle \mathrm{D}|}\Big[\mathds{\hat{1}}_\mathrm{e} -  {i\hat{\Pi}_\mathrm{c} \mu_{\mathrm{BB}}} |\mathrm{B}\rangle\langle \mathrm{B}| + ... \Big]   |\mathrm{D}\rangle\langle \mathrm{B}| \Big[\mathds{\hat{1}}_\mathrm{e} +  {i\hat{\Pi}_\mathrm{c} \mu_{\mathrm{BB}}} |\mathrm{B}\rangle\langle \mathrm{B}| + ... \Big]e^{{i\hat{\Pi}_\mathrm{c} {\mu_{\mathrm{AA}}} |\mathrm{A}\rangle\langle \mathrm{A}|}}\nonumber\\
&= e^{-i\hat{\Pi}_\mathrm{c} \mu_{\mathrm{DD}}|\mathrm{D}\rangle\langle \mathrm{D}|}|\mathrm{D}\rangle\langle \mathrm{B}|e^{i\hat{\Pi}_{\mathrm{c}}\mu_{\mathrm{BB}}}e^{{i\hat{\Pi}_\mathrm{c} {\mu_{\mathrm{AA}}} |\mathrm{A}\rangle\langle \mathrm{A}|}}\nonumber \\
&= e^{i\hat{\Pi}_{\mathrm{c}}\mu_{\mathrm{BB}}}\Big[\mathds{\hat{1}}_\mathrm{e} -  {i\hat{\Pi}_\mathrm{c} \mu_{\mathrm{DD}}} |\mathrm{D}\rangle\langle \mathrm{D}| + ... \Big] |\mathrm{D}\rangle\langle \mathrm{B}|\Big[\mathds{\hat{1}}_\mathrm{e} +  {i\hat{\Pi}_\mathrm{c} \mu_{\mathrm{AA}}} |\mathrm{A}\rangle\langle \mathrm{A}| + ... \Big]\nonumber\\
&=e^{i\hat{\Pi}_{\mathrm{c}}\mu_{\mathrm{BB}}} e^{-i\hat{\Pi}_{\mathrm{c}}\mu_{\mathrm{DD}}}|\mathrm{D}\rangle\langle\mathrm{B}|  = e^{-i\hat{\Pi}_c(\mu_{\mathrm{DD}}-\mu_{\mathrm{BB}})}|\mathrm{D}\rangle\langle \mathrm{B}|
\end{align}
In the second line of Eq.~\ref{eqn:polaron_trans}, we used    $\Big[|\mathrm{A}\rangle\langle \mathrm{A}|,|\mathrm{D}\rangle\langle \mathrm{D}|\Big] = \Big[|\mathrm{B}\rangle\langle \mathrm{B}|,|\mathrm{D}\rangle\langle \mathrm{D}|\Big] = \Big[|\mathrm{B}\rangle\langle \mathrm{B}|,|\mathrm{A}\rangle\langle \mathrm{A}|\Big] = 0$ to split an exponential operator into two parts : $e^{\hat{X} + \hat{Y}} = e^{\hat{X}}e^{\hat{Y}}$ when $[\hat{X},\hat{Y}] = 0$. Following the same procedure, we find:
\begin{equation}\label{sigma-pol}
\hat{U}_\mathrm{pol}^{\dagger}|\mathrm{B}\rangle\langle\mathrm{D}|\hat{U}_\mathrm{pol} = e^{-\frac{i}{\hbar}\sqrt{\frac{2}{\hbar\omega_{\mathrm{c}}^3}} { \chi}(\mu_{\mathrm{BB}}-\mu_{\mathrm{DD}}) \hat{P}_\mathrm{c}}|\mathrm{B}\rangle\langle\mathrm{D}|.
\end{equation}
\begin{equation}\label{AB}
\hat{U}_\mathrm{pol}^{\dagger}|\mathrm{A}\rangle\langle\mathrm{B}|\hat{U}_\mathrm{pol} = e^{-\frac{i}{\hbar}\sqrt{\frac{2}{\hbar\omega_{\mathrm{c}}^3}} { \chi}(\mu_{\mathrm{AA}}-\mu_{\mathrm{BB}}) \hat{P}_\mathrm{c}}|\mathrm{A}\rangle\langle\mathrm{B}|.
\end{equation}
\begin{equation}\label{BA}
\hat{U}_\mathrm{pol}^{\dagger}|\mathrm{B}\rangle\langle\mathrm{A}|\hat{U}_\mathrm{pol} = e^{-\frac{i}{\hbar}\sqrt{\frac{2}{\hbar\omega_{\mathrm{c}}^3}} { \chi}(\mu_{\mathrm{BB}}-\mu_{\mathrm{AA}}) \hat{P}_\mathrm{c}}|\mathrm{B}\rangle\langle\mathrm{A}|.
\end{equation}
\begin{equation}\label{DD}
\hat{U}_\mathrm{pol}^{\dagger}|\mathrm{D}\rangle\langle\mathrm{D}|\hat{U}_\mathrm{pol} = |\mathrm{D}\rangle\langle\mathrm{D}|.
\end{equation}
\begin{equation}\label{BB}
\hat{U}_\mathrm{pol}^{\dagger}|\mathrm{B}\rangle\langle\mathrm{B}|\hat{U}_\mathrm{pol} = |\mathrm{B}\rangle\langle\mathrm{B}|.
\end{equation}
\begin{equation}\label{AA}
\hat{U}_\mathrm{pol}^{\dagger}|\mathrm{A}\rangle\langle\mathrm{A}|\hat{U}_\mathrm{pol} = |\mathrm{A}\rangle\langle\mathrm{A}|.
\end{equation}
Using the result of Eq.~\ref{sigmad-pol} and Eq.~\ref{AA} we find the following polaron transformation
\begin{align}
&\hat{U}_\mathrm{pol}^{\dagger}\big[\hat{Q}_\mathrm{c} + \sqrt{\frac{2}{\hbar\omega_{\mathrm{c}}^3}} \pmb{\hat{\chi}}\cdot \pmb{ \hat{\mu}}  \big]\hat{U}_\mathrm{pol}\nonumber\\
&=\hat{U}_\mathrm{pol}^{\dagger}\hat{Q}_\mathrm{c}\hat{U}_\mathrm{pol} + \sqrt{\frac{2}{\hbar\omega_{\mathrm{c}}^3}} {\chi} \hat{U}_\mathrm{pol}^{\dagger}\Big[\mu_\mathrm{DB}\big[|\mathrm{D}\rangle\langle \mathrm{B}|+|\mathrm{B}\rangle\langle \mathrm{D}|\big]+\mu_\mathrm{BA}\big[|\mathrm{B}\rangle\langle \mathrm{A}|+|\mathrm{A}\rangle\langle \mathrm{B}|\big] \nonumber\\
&~+\mu_\mathrm{DD}|\mathrm{D}\rangle\langle \mathrm{D}| + \mu_\mathrm{BB}|\mathrm{B}\rangle\langle \mathrm{B}| + \mu_\mathrm{AA}|\mathrm{A}\rangle\langle \mathrm{A}|  \Big]\hat{U}_\mathrm{pol}\nonumber\\
&= \hat{Q}_{\mathrm{c}} - \sqrt{\frac{2}{\hbar\omega_{\mathrm{c}}^3}}\chi(\mu_{\mathrm{DD}}|\mathrm{D}\rangle\langle \mathrm{D}| + \mu_{\mathrm{BB}}|\mathrm{B}\rangle\langle \mathrm{B}| + \mu_{\mathrm{AA}}|\mathrm{A}\rangle\langle \mathrm{A}|) + \sqrt{\frac{2}{\hbar\omega_{\mathrm{c}}^3}} \chi \mu_{\mathrm{DB}}\big[e^{-\frac{i}{\hbar}\sqrt{\frac{2}{\hbar\omega_{\mathrm{c}}^3}} { \chi}(\mu_{\mathrm{DD}}-\mu_{\mathrm{BB}}) \hat{P}_\mathrm{c}}|\mathrm{D}\rangle\langle \mathrm{B|} \nonumber\\
&~+ e^{-\frac{i}{\hbar}\sqrt{\frac{2}{\hbar\omega_{\mathrm{c}}^3}} { \chi}(\mu_{\mathrm{BB}}-\mu_{\mathrm{DD}}) \hat{P}_\mathrm{c}}|\mathrm{B}\rangle\langle \mathrm{D}|\big] + \mu_{\mathrm{BA}}\big[e^{-\frac{i}{\hbar}\sqrt{\frac{2}{\hbar\omega_{\mathrm{c}}^3}} { \chi}(\mu_{\mathrm{BB}}-\mu_{\mathrm{AA}}) \hat{P}_\mathrm{c}}|\mathrm{B}\rangle\langle \mathrm{A}| + e^{-\frac{i}{\hbar}\sqrt{\frac{2}{\hbar\omega_{\mathrm{c}}^3}} { \chi}(\mu_{\mathrm{AA}}-\mu_{\mathrm{BB}}) \hat{P}_\mathrm{c}}|\mathrm{A}\rangle\langle \mathrm{B}| \big]\nonumber\\
&~+\sqrt{\frac{2}{\hbar\omega_{\mathrm{c}}^3}}\chi(\mu_{\mathrm{DD}}|\mathrm{D}\rangle\langle \mathrm{D}| + \mu_{\mathrm{BB}}|\mathrm{B}\rangle\langle \mathrm{B}| + \mu_{\mathrm{AA}}|\mathrm{A}\rangle\langle \mathrm{A}|)\nonumber\\
&= \hat{Q}_{\mathrm{c}} + \sqrt{\frac{2}{\hbar\omega_{\mathrm{c}}^3}} \chi \mu_{\mathrm{DB}}\bigg[e^{-\frac{i}{\hbar}\sqrt{\frac{2}{\hbar\omega_{\mathrm{c}}^3}} { \chi}(\mu_{\mathrm{DD}}-\mu_{\mathrm{BB}}) \hat{P}_\mathrm{c}}|\mathrm{D}\rangle\langle \mathrm{B|} + e^{-\frac{i}{\hbar}\sqrt{\frac{2}{\hbar\omega_{\mathrm{c}}^3}} { \chi}(\mu_{\mathrm{BB}}-\mu_{\mathrm{DD}}) \hat{P}_\mathrm{c}}|\mathrm{B}\rangle\langle \mathrm{D}|\bigg]\nonumber\\
&~+ \sqrt{\frac{2}{\hbar\omega_{\mathrm{c}}^3}} { \chi}\mu_{\mathrm{BA}}\bigg[e^{-\frac{i}{\hbar}\sqrt{\frac{2}{\hbar\omega_{\mathrm{c}}^3}} { \chi}(\mu_{\mathrm{BB}}-\mu_{\mathrm{AA}}) \hat{P}_\mathrm{c}}|\mathrm{B}\rangle\langle \mathrm{A}| + e^{-\frac{i}{\hbar}\sqrt{\frac{2}{\hbar\omega_{\mathrm{c}}^3}} { \chi}(\mu_{\mathrm{AA}}-\mu_{\mathrm{BB}}) \hat{P}_\mathrm{c}}|\mathrm{A}\rangle\langle \mathrm{B}|\bigg]
\end{align}
With the above results, and the fact that $e^{\hat{Y}}\hat{O}(\hat{X})e^{-\hat{Y}}=\hat{O}(e^{\hat{Y}}\hat{X}e^{-\hat{Y}})$ for a unitary operator $e^{\hat{Y}}$, we apply $\hat{U}_\mathrm{pol}$ to $\hat{H}_{\mathrm{PF}}$ and find the following polaron transformed PF Hamiltonian:
\begin{align}\label{eq:polaron_trans_2}
\tilde{H}_\mathrm{PF}&=\hat{U}_\mathrm{pol}^{\dagger}\hat{H}_{\mathrm{PF}}\hat{U}_\mathrm{pol} =\hat{U}_\mathrm{pol}^\dagger  \hat{H}_\mathrm{m}  \hat{U}_\mathrm{pol} +\frac{1}{2}\hat{P}_\mathrm{c}^{2}+\hat{U}_\mathrm{pol}^{\dagger}\frac{1}{2}\omega_{\mathrm{c}}^{2}\big[\hat{Q}_\mathrm{c} + \sqrt{\frac{2}{\hbar\omega_{\mathrm{c}}^3}} \pmb{\hat{\chi}}\cdot \pmb{ \hat{\mu}}  \big]^2 \hat{U}_\mathrm{pol} \nonumber\\
&=  \hat{U}_\mathrm{pol}^\dagger  \hat{H}_\mathrm{m}  \hat{U}_\mathrm{pol} +\frac{1}{2}\hat{P}_\mathrm{c}^{2}+\frac{1}{2}\omega_{\mathrm{c}}^{2}\Bigg[\hat{Q}_{\mathrm{c}} + \sqrt{\frac{2}{\hbar\omega_{\mathrm{c}}^3}} \chi \mu_{\mathrm{DB}}\big[e^{-\frac{i}{\hbar}\sqrt{\frac{2}{\hbar\omega_{\mathrm{c}}^3}} { \chi}(\mu_{\mathrm{DD}}-\mu_{\mathrm{BB}}) \hat{P}_\mathrm{c}}|\mathrm{D}\rangle\langle \mathrm{B|} \nonumber\\
&~~~+ e^{-\frac{i}{\hbar}\sqrt{\frac{2}{\hbar\omega_{\mathrm{c}}^3}} { \chi}(\mu_{\mathrm{BB}}-\mu_{\mathrm{DD}}) \hat{P}_\mathrm{c}}|\mathrm{B}\rangle\langle \mathrm{D}|\big]+ \sqrt{\frac{2}{\hbar\omega_{\mathrm{c}}^3}} \chi\mu_{\mathrm{BA}}\big[e^{-\frac{i}{\hbar}\sqrt{\frac{2}{\hbar\omega_{\mathrm{c}}^3}} { \chi}(\mu_{\mathrm{BB}}-\mu_{\mathrm{AA}}) \hat{P}_\mathrm{c}}|\mathrm{B}\rangle\langle \mathrm{A}| \nonumber\\
&~~~+ e^{-\frac{i}{\hbar}\sqrt{\frac{2}{\hbar\omega_{\mathrm{c}}^3}} { \chi}(\mu_{\mathrm{AA}}-\mu_{\mathrm{BB}}) \hat{P}_\mathrm{c}}|\mathrm{A}\rangle\langle \mathrm{B}|  \big]\Bigg]^2  \nonumber \\
&=  \hat{U}_\mathrm{pol}^\dagger  \hat{H}_\mathrm{m}  \hat{U}_\mathrm{pol} + \frac{1}{2}\hat{P}_{\mathrm{c}}^2 + \frac{1}{2}\omega_{\mathrm{c}}^2\hat{Q}_{\mathrm{c}}^2 + \frac{1}{\hbar\omega_{\mathrm{c}}}\chi^2\mu_{\mathrm{DB}}^2\big[|\mathrm{D}\rangle\langle\mathrm{D}| + |\mathrm{B}\rangle\langle\mathrm{B}|\big] + \frac{1}{\hbar\omega_{\mathrm{c}}}\chi^2\mu_{\mathrm{BA}}^2\big[|\mathrm{B}\rangle\langle\mathrm{B}| \nonumber\\
&~~~+ |\mathrm{A}\rangle\langle\mathrm{A}|\big] + \omega_{\mathrm{c}}^2\hat{Q}_{\mathrm{c}} \sqrt{\frac{2}{\hbar\omega_{\mathrm{c}}^3}}\chi \mu_{\mathrm{DB}}\big[e^{-\frac{i}{\hbar}\sqrt{\frac{2}{\hbar\omega_{\mathrm{c}}^3}} { \chi}(\mu_{\mathrm{DD}}-\mu_{\mathrm{BB}}) \hat{P}_\mathrm{c}}|\mathrm{D}\rangle\langle \mathrm{B|}
+e^{-\frac{i}{\hbar}\sqrt{\frac{2}{\hbar\omega_{\mathrm{c}}^3}} { \chi}(\mu_{\mathrm{BB}}-\mu_{\mathrm{DD}}) \hat{P}_\mathrm{c}}|\mathrm{B}\rangle\langle \mathrm{D}|\big]\nonumber\\
&~~~+ \frac{1}{\hbar \omega_{\mathrm{c}}}\chi^2\mu_{\mathrm{DB}}\mu_{\mathrm{BA}}\big[e^{-\frac{i}{\hbar}\sqrt{\frac{2}{\hbar\omega_{\mathrm{c}}^3}} { \chi}(\mu_{\mathrm{DD}}-\mu_{\mathrm{AA}}) \hat{P}_\mathrm{c}}|\mathrm{D}\rangle\langle \mathrm{A}| + e^{-\frac{i}{\hbar}\sqrt{\frac{2}{\hbar\omega_{\mathrm{c}}^3}} { \chi}(\mu_{\mathrm{AA}}-\mu_{\mathrm{DD}}) \hat{P}_\mathrm{c}}|\mathrm{A}\rangle\langle \mathrm{D}|\big] \nonumber\\
&~~~+ \omega_{\mathrm{c}}^2\hat{Q}_{\mathrm{c}} \sqrt{\frac{2}{\hbar\omega_{\mathrm{c}}^3}}\chi \mu_{\mathrm{BA}}\big[e^{-\frac{i}{\hbar}\sqrt{\frac{2}{\hbar\omega_{\mathrm{c}}^3}} { \chi}(\mu_{\mathrm{BB}}-\mu_{\mathrm{AA}}) \hat{P}_\mathrm{c}}|\mathrm{B}\rangle\langle \mathrm{A}|+
e^{-\frac{i}{\hbar}\sqrt{\frac{2}{\hbar\omega_{\mathrm{c}}^3}} { \chi}(\mu_{\mathrm{AA}}-\mu_{\mathrm{BB}}) \hat{P}_\mathrm{c}}|\mathrm{A}\rangle\langle \mathrm{B}|\big]\nonumber\\
&= \hat{U}_{\mathrm{pol}}^{
\dagger} \hat{H}_{\mathrm{m}} \hat{U}_{\mathrm{pol}} + (
\hat{a}^{\dagger}\hat{a} + \frac{1}{2})\hbar \omega_{\mathrm{c}} + \frac{\hbar g_{\mathrm{c}}^2}{\omega_{\mathrm{c}}}\big[|\mathrm{D}\rangle\langle \mathrm{D}|+|\mathrm{B}\rangle\langle \mathrm{B}|\big] + \frac{\hbar \eta_{\mathrm{c}}^2}{\omega_{\mathrm{c}}}\big[|\mathrm{B}\rangle\langle \mathrm{B}|+|\mathrm{A}\rangle\langle \mathrm{A}|\big] \nonumber\\
&~~~+ \hbar g_{\mathrm{c}} (\hat{a}+\hat{a}^{\dagger})\big[e^{-\frac{i}{\hbar}\sqrt{\frac{2}{\hbar\omega_{\mathrm{c}}^3}} { \chi}(\mu_{\mathrm{DD}}-\mu_{\mathrm{BB}}) \hat{P}_\mathrm{c}}|\mathrm{D}\rangle\langle \mathrm{B}| + e^{-\frac{i}{\hbar}\sqrt{\frac{2}{\hbar\omega_{\mathrm{c}}^3}} { \chi}(\mu_{\mathrm{BB}}-\mu_{\mathrm{DD}}) \hat{P}_\mathrm{c}}|\mathrm{B}\rangle\langle \mathrm{D}|\big]\nonumber\\
&~~~+ \frac{\hbar g_{\mathrm{c}} \eta_{\mathrm{c}}}{\omega_{\mathrm{c}}}\big[e^{-\frac{i}{\hbar}\sqrt{\frac{2}{\hbar\omega_{\mathrm{c}}^3}} { \chi}(\mu_{\mathrm{DD}}-\mu_{\mathrm{AA}}) \hat{P}_\mathrm{c}}|\mathrm{D}\rangle\langle \mathrm{A}| + e^{-\frac{i}{\hbar}\sqrt{\frac{2}{\hbar\omega_{\mathrm{c}}^3}} { \chi}(\mu_{\mathrm{AA}}-\mu_{\mathrm{DD}}) \hat{P}_\mathrm{c}}|\mathrm{A}\rangle\langle \mathrm{D}|\big] \nonumber\\
&~~~+ \hbar \eta_{\mathrm{c}} (\hat{a}+\hat{a}^{\dagger})\big[e^{-\frac{i}{\hbar}\sqrt{\frac{2}{\hbar\omega_{\mathrm{c}}^3}} { \chi}(\mu_{\mathrm{BB}}-\mu_{\mathrm{AA}}) \hat{P}_\mathrm{c}}|\mathrm{B}\rangle\langle \mathrm{A}| +  e^{-\frac{i}{\hbar}\sqrt{\frac{2}{\hbar\omega_{\mathrm{c}}^3}} { \chi}(\mu_{\mathrm{AA}}-\mu_{\mathrm{BB}}) \hat{P}_\mathrm{c}}|\mathrm{A}\rangle\langle \mathrm{B}|\big]
\end{align}
From the second to third line of Eq.~\ref{eq:polaron_trans_2}, we used the fact that, $(\hat{a}+\hat{b}+\hat{c})^2 = (\hat{a}+\hat{b}+\hat{c})\cdot (\hat{a}+\hat{b}+\hat{c})$, where,
\begin{align}
\hat{a}&= \hat{Q}_{\mathrm{c}},\nonumber\\ \hat{b}&=\sqrt{\frac{2}{\hbar\omega_{\mathrm{c}}^3}} \chi \mu_{\mathrm{DB}}\big[e^{-\frac{i}{\hbar}\sqrt{\frac{2}{\hbar\omega_{\mathrm{c}}^3}} {\chi}(\mu_{\mathrm{DD}}-\mu_{\mathrm{BB}}) \hat{P}_\mathrm{c}}|\mathrm{D}\rangle\langle \mathrm{B|} + e^{-\frac{i}{\hbar}\sqrt{\frac{2}{\hbar\omega_{\mathrm{c}}^3}} { \chi}(\mu_{\mathrm{BB}}-\mu_{\mathrm{DD}}) \hat{P}_\mathrm{c}}|\mathrm{B}\rangle\langle \mathrm{D}|\big],\nonumber\\
\hat{c}&= \sqrt{\frac{2}{\hbar\omega_{\mathrm{c}}^3}} \chi \mu_{\mathrm{BA}}\big[e^{-\frac{i}{\hbar}\sqrt{\frac{2}{\hbar\omega_{\mathrm{c}}^3}} {\chi}(\mu_{\mathrm{BB}}-\mu_{\mathrm{AA}})\hat{P}_\mathrm{c}}|\mathrm{B}\rangle\langle \mathrm{A}| + e^{-\frac{i}{\hbar}\sqrt{\frac{2}{\hbar\omega_{\mathrm{c}}^3}} {\chi}(\mu_{\mathrm{AA}}-\mu_{\mathrm{BB}}) \hat{P}_\mathrm{c}}|\mathrm{A}\rangle\langle \mathrm{B}|\big].
\end{align}
The fourth line of Eq.~\ref{eq:polaron_trans_2} is the final expression for the polaron transformed Hamiltonian of a DBA system where the cavity frequency is not in resonance with electronic transition. Note that, to proceed from the third to fourth line of Eq.~\ref{eq:polaron_trans_2} we used $\hat{Q}_{\mathrm{c}} = \sqrt{\frac{\hbar}{2\omega_{\mathrm{c}}}}(\hat{a}+\hat{a}^{\dagger})$, $\hbar g_{\mathrm{c}} = \chi \mu_{\mathrm{DB}}$, and $\hbar \eta_{\mathrm{c}} = \chi \mu_{\mathrm{BA}}$. Thus $\hat{U}_{\mathrm{pol}}^{\dagger}\hat{H}_{\mathrm{m}}\hat{U}_{\mathrm{pol}}$ is
\begin{align}
\hat{U}_\mathrm{pol}^{\dagger}\hat{H}_\mathrm{m}\hat{U}_\mathrm{pol} =& \frac{\hat{P}_{\mathrm{s}}^2}{2M_{\mathrm{s}}} +\sum_{i} \frac{1}{2}M_{\mathrm{s}}\omega_{\mathrm{s}}^2(R_{\mathrm{s}}-R_{i}^{0})^2|i\rangle\langle i| + \sum_{i} U_{i}|i\rangle\langle i|\\
&+V_{\mathrm{DB}}(e^{-\frac{i}{\hbar}\sqrt{\frac{2}{\hbar\omega_{\mathrm{c}}^3}} {\chi}(\mu_{\mathrm{DD}}-\mu_{\mathrm{BB}}) \hat{P}_\mathrm{c}}|\mathrm{D}\rangle\langle\mathrm{B}|+e^{-\frac{i}{\hbar}\sqrt{\frac{2}{\hbar\omega_{\mathrm{c}}^3}} {\chi}(\mu_{\mathrm{BB}}-\mu_{\mathrm{DD}}) \hat{P}_\mathrm{c}}|\mathrm{B}\rangle\langle\mathrm{D}|) \nonumber\\
&+V_{\mathrm{BA}}(e^{-\frac{i}{\hbar}\sqrt{\frac{2}{\hbar\omega_{\mathrm{c}}^3}} {\chi}(\mu_{\mathrm{BB}}-\mu_{\mathrm{AA}}) \hat{P}_\mathrm{c}}|\mathrm{B}\rangle\langle\mathrm{A}|+e^{-\frac{i}{\hbar}\sqrt{\frac{2}{\hbar\omega_{\mathrm{c}}^3}} {\chi}(\mu_{\mathrm{AA}}-\mu_{\mathrm{BB}}) \hat{P}_\mathrm{c}}|\mathrm{A}\rangle\langle\mathrm{B}|) + \hat{H}_{\mathrm{sb}} \nonumber
\end{align}
With this polaron transformed Hamiltonian in Eq.~\ref{eq:polaron_trans_2}, we can extract the effective couplings between the donor and bridge ($\tilde{V}_{n,l}^{\mathrm{DB}}$), and the bridge and acceptor($\tilde{V}_{l,m}^{\mathrm{BA}}$) state as follows
\begin{align}\label{eq.V_DB}
\tilde{V}_{n,l}^{\mathrm{DB}} &=  \langle \mathrm{D},n | \tilde{H}_{\mathrm{PF}}-\hat{T}_{\mathrm{s}} - \hat{H}_{\mathrm{sb}}|\mathrm{B},l\rangle \nonumber\\
&= \langle \mathrm{D},n|V_{\mathrm{DB}}e^{-\frac{i}{
\hbar}\sqrt{\frac{2}{\hbar\omega_\mathrm{c}^3}}\chi (\mu_{\mathrm{DD}}-\mu_{\mathrm{BB}})\hat{P}_{\mathrm{c}}}|\mathrm{D}\rangle \langle \mathrm{B}|\mathrm{B},l\rangle \nonumber\\ 
&~+ \hbar g_{\mathrm{c}}\langle \mathrm{D},n | (\hat{a} + \hat{a}^{\dagger})e^{-\frac{i}{
\hbar}\sqrt{\frac{2}{\hbar\omega_\mathrm{c}^3}}\chi (\mu_{\mathrm{DD}}-\mu_{\mathrm{BB}})\hat{P}_{\mathrm{c}}}|\mathrm{D}\rangle\langle \mathrm{B}|\mathrm{B},l\rangle \nonumber\\
& = V_{\mathrm{DB}}S_{n,l}^{\mathrm{DB}} + \hbar g_{\mathrm{c}}\big[\sqrt{n}S_{n-1,l}^{\mathrm{DB}} + \sqrt{n+1}S_{n+1,l}^{\mathrm{DB}}\big]
\end{align}
Similarly, 
\begin{align}\label{eq.V_BA}
\tilde{V}_{l,m}^{\mathrm{BA}} &= \langle \mathrm{B},l|\tilde{H}_{\mathrm{PF}}-\hat{T}_{\mathrm{s}} - \hat{H}_{\mathrm{sb}}|\mathrm{A},m\rangle\nonumber\\
&=V_{\mathrm{BA}}S_{l,m}^{\mathrm{BA}} + \hbar\eta_{\mathrm{c}}\big[\sqrt{n}S_{l-1,m}^{\mathrm{BA}}+\sqrt{l+1}S_{l+1,m}^{\mathrm{BA}}\big]
\end{align}
The direct donor to acceptor coupling is found to be:
\begin{align}\label{eq.V_DA}
\tilde{V}_{n,m}^{\mathrm{DA}} &= \langle \mathrm{D},n|\tilde{H}_{\mathrm{PF}} - \hat{T}_{\mathrm{S}} - \hat{H}_{\mathrm{sb}} | \mathrm{A},m\rangle \nonumber\\
& = \frac{\hbar g_{\mathrm{c}} \eta_{\mathrm{c}}}{\omega_{\mathrm{c}}}S_{n,m}^{\mathrm{DA}}    
\end{align}
where, $S_{n,l}^{\mathrm{DB}} = \langle n|e^{-i/\hbar \sqrt{2/\hbar \omega_{\mathrm{c}}^3}\chi(\mu_{\mathrm{DD}}-\mu_{\mathrm{BB}})\hat{P}_{\mathrm{c}}}|l\rangle$, $S_{l,m}^{\mathrm{BA}} = \langle l|e^{-i/\hbar \sqrt{2/\hbar \omega_{\mathrm{c}}^3}\chi(\mu_{\mathrm{BB}}-\mu_{\mathrm{AA}})\hat{P}_{\mathrm{c}}}|m\rangle$, and $S_{n,m}^{\mathrm{DA}} = \langle n|e^{-i/\hbar \sqrt{2/\hbar \omega_{\mathrm{c}}^3}\chi(\mu_{\mathrm{DD}}-\mu_{\mathrm{AA}})\hat{P}_{\mathrm{c}}}|m\rangle$

The bridge mediated effective donor-acceptor coupling ($V_{n,m}^{'\mathrm{DA}}$)  can be calculated by perturbative technique i.e. the transformation method~\cite{fleming1975}, and  
\begin{align}\label{eq.V_DA_indirect}
\tilde{V}_{n,m}^{'\mathrm{DA}} =  -\sum_l \frac{\tilde{V}_{nl}^{*\mathrm{DB}}\tilde{V}_{l,m}^{\mathrm{BA}}}{2}\bigg[\frac{1}{(U_{\mathrm{B}}-U_{\mathrm{A}})+(l-m)\hbar\omega_{\mathrm{c}}} + \frac{1}{(U_{\mathrm{B}}-U_{\mathrm{D}})+(l-n)\hbar\omega_{\mathrm{c}}}\bigg]   
\end{align}
Hence, the total $\mathrm{DA}$ coupling is:
\begin{align}\label{eq.total_DA_coup}
F_{n,m}^{\mathrm{DA}} &= \tilde{V}_{n,m}^{\mathrm{DA}} + \tilde{V}_{n,m}^{'\mathrm{DA}}\nonumber\\
&= \frac{\hbar g_\mathrm{c}\eta_{\mathrm{c}}}{\omega_{\mathrm{c}}}S_{n,m}^{\mathrm{DA}}-\sum_l \frac{\tilde{V}_{n,l}^{\mathrm{DB}}\tilde{V}_{l,m}^{\mathrm{BA}}}{2}\bigg[\frac{1}{(U_{\mathrm{B}}-U_{\mathrm{A}})+(l-m)\hbar\omega_{\mathrm{c}}} + \frac{1}{(U_{\mathrm{B}}-U_{\mathrm{D}})+(l-n)\hbar\omega_{\mathrm{c}}}\bigg]
\end{align}
With the above polaron transformation, the polariton mediated electron transfer (PMET) rate is: 
\begin{eqnarray}\label{eqn:rate_each_channel}
k_\mathrm{FGR} = \sum_{n}{P}_n \sum_{m}\frac{|F_{n,m}^{\mathrm{DA}}|^2}{\hbar}
\sqrt{\frac{\pi}{\lambda_{\mathrm{DA}}k_BT}}\mathrm{exp}\bigg[-\frac{(\Delta G_{n,m} + \lambda_{\mathrm{DA}} )^2}{4\lambda_{\mathrm{DA}}k_BT}\bigg],
\end{eqnarray}
where, $\Delta G_{n,m}$ is the effective driving force for each channel
\begin{eqnarray}\label{eq:driving_force_nm}
\Delta G_{n,m} &=& -(U_{\mathrm{D}}-U_{\mathrm{A}}) + (m-n)\hbar \omega_c - \sum_l \frac{(\tilde{V}_{l,n}^{\mathrm{BA}})^{\mathrm{T}}\tilde{V}_{l,m}^{\mathrm{BA}}}{(U_{\mathrm{B}}-U_{\mathrm{A}}) + (l-m)\hbar\omega_c} \nonumber\\
&&~+ \sum_l \frac{(\tilde{V}_{n,l}^{\mathrm{DB}})(\tilde{V}_{m,l}^{\mathrm{DB}})^{\mathrm{T}}}{(U_{\mathrm{B}}-U_{\mathrm{D}}) + (l-n)\hbar\omega_c}
\end{eqnarray}
with,
\begin{equation}
P_n= \frac{\mathrm{exp}[-\beta n \hbar \omega_c]}{\sum_n \mathrm{exp}[-\beta n \hbar \omega_c]}.
\end{equation}
\section{\normalsize Derivation of the PMET Rate for the On-Resonance Case}
As in the previous section, we introduce the following polaron transformation of the photonic DOF
\begin{equation}
\hat{U}_\mathrm{pol}= \mathrm{exp}\Bigg[{\frac{i}{\hbar}\sqrt{2\over \hbar\omega_\mathrm{c}^3}{\chi} \sum_{j\in{\mathrm{D,A}}}\mu_{jj}|j \rangle \langle j|\hat{P}_\mathrm{c}}\Bigg].  
\end{equation}
In this system the bridge state is not coupled with the light field. Following the same derivation as described in the previous section, we find:
\begin{align}\label{eq:pol_tran_iv_ct}
\tilde{H}_\mathrm{PF}&=\hat{U}_\mathrm{pol}^{\dagger}\hat{H}_{\mathrm{PF}}\hat{U}_\mathrm{pol} =\hat{U}_\mathrm{pol}^\dagger  \hat{H}_\mathrm{m}  \hat{U}_\mathrm{pol} +\frac{1}{2}\hat{P}_\mathrm{c}^{2}+\hat{U}_\mathrm{pol}^{\dagger}\frac{1}{2}\omega_{\mathrm{c}}^{2}\big[\hat{Q}_\mathrm{c} + \sqrt{\frac{2}{\hbar\omega_{\mathrm{c}}^3}} \pmb{\hat{\chi}}\cdot \pmb{\hat{\mu}}  \big]^2 \hat{U}_\mathrm{pol} \\
&=  \hat{U}_\mathrm{pol}^\dagger  \hat{H}_\mathrm{m}  \hat{U}_\mathrm{pol} +\frac{1}{2}\hat{P}_\mathrm{c}^{2}+\frac{1}{2}\omega_{\mathrm{c}}^{2}\big[\hat{Q}_\mathrm{c} + \sqrt{\frac{2}{\hbar\omega_{\mathrm{c}}^3}} \chi \mu_\mathrm{DA}(e^{\frac{i}{\hbar}\sqrt{\frac{2}{\hbar\omega_{\mathrm{c}}^3}}  \chi\Delta\mu \hat{P}_\mathrm{c}}\hat{\sigma}^\dagger+e^{-\frac{i}{\hbar}\sqrt{\frac{2}{\hbar\omega_{\mathrm{c}}^3}}  \chi\Delta\mu \hat{P}_\mathrm{c}}\hat{\sigma})  \big]^2  \nonumber \\
&=  \hat{U}_\mathrm{pol}^\dagger  \hat{H}_\mathrm{m}  \hat{U}_\mathrm{pol} + (\hat{a}^{\dagger} \hat{a} + {1\over 2}) \hbar\omega_{\mathrm{c}} + \hbar g_\mathrm{c} (\hat{a}^{\dagger} + \hat{a}) (e^{\frac{i}{\hbar}\sqrt{\frac{2}{\hbar\omega_{\mathrm{c}}^3}} \chi\Delta\mu \hat{P}_\mathrm{c}}\hat{\sigma}^\dagger+e^{-\frac{i}{\hbar}\sqrt{\frac{2}{\hbar\omega_{\mathrm{c}}^3}} \chi\Delta\mu \hat{P}_\mathrm{c}}\hat{\sigma}) +  {\hbar g_\mathrm{c}^2\over {\omega_{\mathrm{c}}}}\hat{\mathds{1}}_\mathrm{e}. \nonumber
\end{align}
where $\Delta \mu = \mu_{\mathrm{DD}}-\mu_{\mathrm{AA}}$, $\hat{\sigma}^\dagger = |\mathrm{A}\rangle\langle \mathrm{D}|$, $\hat{\sigma} = |\mathrm{D}\rangle\langle \mathrm{A}|$, $\hat{\mathds{1}}_\mathrm{e}=|\mathrm{D}\rangle\langle\mathrm{D}|+|\mathrm{A}\rangle\langle\mathrm{A}|$, and $\hbar g_c = \chi\mu_{\mathrm{DA}}$. Finally, $\hat{U}_{\mathrm{pol}}^{\dagger}\hat{H}_{\mathrm{m}}\hat{U}_{\mathrm{pol}}$ is
\begin{align}
\hat{U}_\mathrm{pol}^{\dagger}\hat{H}_\mathrm{m}\hat{U}_\mathrm{pol} =& \frac{\hat{P}_{\mathrm{s}}^2}{2M_{\mathrm{s}}} +\sum_{i} \frac{1}{2}M_{\mathrm{s}}\omega_{\mathrm{s}}^2(R_{\mathrm{s}}-R_{i}^{0})^2|i\rangle\langle i| + \sum_{i} U_{i}|i\rangle\langle i|\\
&+V_{\mathrm{DB}}(e^{-\frac{i}{\hbar}\sqrt{\frac{2}{\hbar\omega_{\mathrm{c}}^3}} {\chi}\mu_{\mathrm{DD}} \hat{P}_\mathrm{c}}|\mathrm{D}\rangle\langle\mathrm{B}|+e^{\frac{i}{\hbar}\sqrt{\frac{2}{\hbar\omega_{\mathrm{c}}^3}} {\chi}\mu_{\mathrm{DD}} \hat{P}_\mathrm{c}}|\mathrm{B}\rangle\langle\mathrm{D}|) \nonumber\\
&+V_{\mathrm{BA}}(e^{\frac{i}{\hbar}\sqrt{\frac{2}{\hbar\omega_{\mathrm{c}}^3}} {\chi}\mu_{\mathrm{AA}} \hat{P}_\mathrm{c}}|\mathrm{B}\rangle\langle\mathrm{A}|+e^{-\frac{i}{\hbar}\sqrt{\frac{2}{\hbar\omega_{\mathrm{c}}^3}} {\chi}\mu_{\mathrm{AA}} \hat{P}_\mathrm{c}}|\mathrm{A}\rangle\langle\mathrm{B}|) + \hat{H}_{\mathrm{sb}} \nonumber
\end{align}
With the polaron transformed Hamiltonian in Eq.~\ref{eq:pol_tran_iv_ct}, we can compute the effective couplings between the donor and bridge ($\tilde{V}_{n,l}^{\mathrm{DB}}$), and the bridge and acceptor($\tilde{V}_{l,m}^{\mathrm{BA}}$) states as follows:
\begin{align}\label{eq.V_DB_ivct}
\tilde{V}_{n,0}^{\mathrm{DB}} &=  \langle \mathrm{D},n | \tilde{H}_{\mathrm{PF}}-\hat{T}_{\mathrm{s}} - \hat{H}_{\mathrm{sb}}|\mathrm{B},0\rangle \nonumber\\
&= \langle \mathrm{D},n|V_{\mathrm{DB}}e^{-\frac{i}{
\hbar}\sqrt{\frac{2}{\hbar\omega_\mathrm{c}^3}}\chi \mu_{\mathrm{DD}}\hat{P}_{\mathrm{c}}}|\mathrm{D}\rangle \langle \mathrm{B}|\mathrm{B},0\rangle\nonumber\\ 
& = V_{\mathrm{DB}}S_{n,0}^{\mathrm{DB}}
\end{align}
Similarly, 
\begin{align}\label{eq.V_BA_ivct}
\tilde{V}_{0,m}^{\mathrm{BA}} &= \langle \mathrm{B},0|\tilde{H}_{\mathrm{PF}}-\hat{T}_{\mathrm{s}} - \hat{H}_{\mathrm{sb}}|\mathrm{A},m\rangle\nonumber\\
&=V_{\mathrm{BA}}S_{0,m}^{\mathrm{BA}}
\end{align}
The direct donor to acceptor coupling is:
\begin{align}\label{eq.V_DA_ivct}
\tilde{V}_{nm}^{\mathrm{DA}} &= \langle \mathrm{D},n|\tilde{H}_{\mathrm{PF}} - \hat{T}_{\mathrm{S}} - \hat{H}_{\mathrm{sb}} | \mathrm{A},m\rangle \nonumber\\
& = \hbar g_{c}[\sqrt{n}S_{n-1,m} + \sqrt{n+1}S_{n+1,m}]
\end{align}
where, $S_{n,0}^{\mathrm{DB}} = \langle n|e^{-i/\hbar \sqrt{2/\hbar \omega_{\mathrm{c}}^3}\chi\mu_{\mathrm{DD}}\hat{P}_{\mathrm{c}}}|0\rangle$, $S_{0,m}^{\mathrm{BA}} = \langle 0|e^{i/\hbar \sqrt{2/\hbar \omega_{\mathrm{c}}^3}\chi\mu_{\mathrm{AA}}\hat{P}_{\mathrm{c}}}|m\rangle$, and  $S_{n,m}^{\mathrm{DA}} = \langle n|e^{-i/\hbar \sqrt{2/\hbar \omega_{\mathrm{c}}^3}\chi(\mu_{\mathrm{DD}}-\mu_{\mathrm{AA}})\hat{P}_{\mathrm{c}}}|m\rangle$

We again calculate the through-bridge interactions using the transformation method.~\cite{fleming1975} The indirect coupling between donor and acceptor state is:
\begin{align}\label{eq.V_DA_indirect_ivct}
\tilde{V}_{n,m}^{'\mathrm{DA}} = - \frac{\tilde{V}_{n0}^{*\mathrm{DB}}\tilde{V}_{0m}^{\mathrm{BA}}}{2}\bigg[\frac{1}{(U_{\mathrm{B}}-U_{\mathrm{A}})-m\hbar\omega_{\mathrm{c}}} + \frac{1}{(U_{\mathrm{B}}-U_{\mathrm{D}})-n\hbar\omega_{\mathrm{c}}}\bigg]   
\end{align}
Hence, the total donor to acceptor coupling is: \begin{align}\label{eq.total_DA_coup_ivct}
F_{n,m}^{\mathrm{DA}} &= \tilde{V}_{n,m}^{\mathrm{DA}} + \tilde{V}_{n,m}^{'\mathrm{DA}}\nonumber\\
&= \hbar g_{c}[\sqrt{n}S_{n-1,m} + \sqrt{n+1}S_{n+1,m}] -\frac{\tilde{V}_{n,0}^{\mathrm{DB}}\tilde{V}_{0,m}^{\mathrm{BA}}}{2}\bigg[\frac{1}{(U_{\mathrm{B}}-U_{\mathrm{A}})-m\hbar\omega_{\mathrm{c}}} + \frac{1}{(U_{\mathrm{B}}-U_{\mathrm{D}})-n\hbar\omega_{\mathrm{c}}}\bigg],
\end{align}
which is found in Eq.~(7) of the main text. Finally, the polariton mediated electron transfer (PMET) rate is: 
\begin{eqnarray}\label{eqn:rate_each_channel_ivct}
k_\mathrm{FGR} = \sum_{n}{P}_n \sum_{m}\frac{|V_{nm}^{\mathrm{DA}}|^2}{\hbar}
\sqrt{\frac{\pi}{\lambda_{\mathrm{DA}}k_BT}}\mathrm{exp}\bigg[-\frac{(\Delta G_{nm} + \lambda_{\mathrm{DA}} )^2}{4\lambda_{\mathrm{DA}}k_BT}\bigg],
\end{eqnarray}
where, $\Delta G_{nm}$ is the effective driving force for each channel
\begin{eqnarray}\label{eq:driving_force_nm_ivct}
\Delta G_{nm} &=& -(U_{\mathrm{D}}-U_{\mathrm{A}}) + (m-n)\hbar \omega_c - \frac{(\tilde{V}_{0,n}^{\mathrm{BA}})^{\mathrm{T}}\tilde{V}_{0,m}^{\mathrm{BA}}}{(U_{\mathrm{B}}-U_{\mathrm{A}})-m\hbar\omega_c} + \frac{\tilde{V}_{n,0}^{\mathrm{DB}}(\tilde{V}_{m,0}^{\mathrm{DB}})^{\mathrm{T}}}{(U_{\mathrm{B}}-U_{\mathrm{D}})-n\hbar\omega_c}~~
\end{eqnarray}
with,
\begin{equation}
P_n= \frac{\mathrm{exp}[-\beta n \hbar \omega_c]}{\sum_n \mathrm{exp}[-\beta n \hbar \omega_c]}.
\end{equation}
\section{\normalsize Additional Numerical Results}
\begin{figure}
 \centering
  \begin{minipage}[t]{1.0\linewidth}
     \centering
     \includegraphics[width=\linewidth]{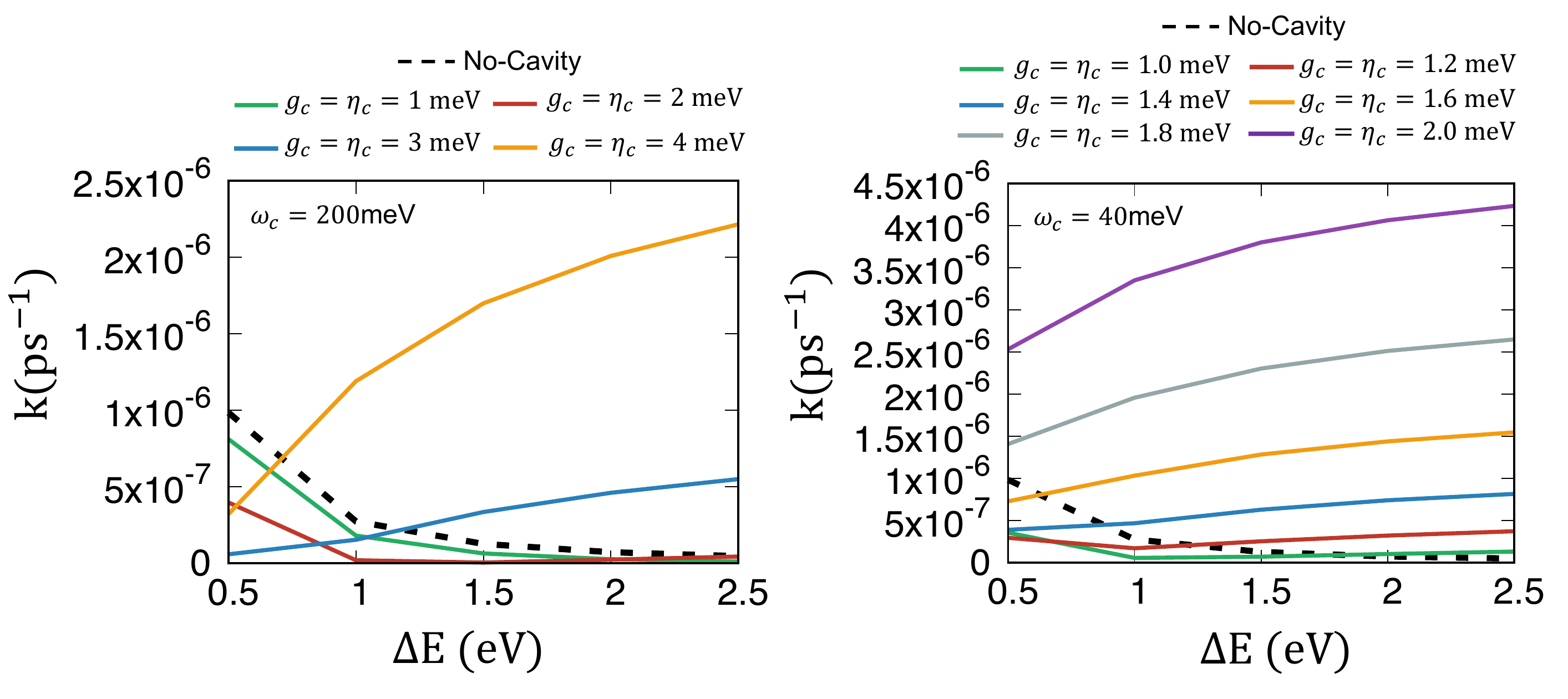}
  \end{minipage}%
   \caption{PMET rate constant obtained over a range of $\Delta E=U_{\mathrm{B}}-U_{\mathrm{D}}$ for off resonance condition with (A) $\hbar \omega_c$ = 200 meV and (B) $\hbar \omega_c$ = 40 meV. Different colored lines correspond to different light-matter interaction strength.}
\label{fig:delE_plot}
\end{figure}

Here, we present additional numerical results for polariton mediated electron transfer (PMET) in the off-resonance limit. We describe the PMET rate obtained from the FGR rate expression (Eq.~\ref{eqn:rate_each_channel}). Figure~\ref{fig:delE_plot}(A) and \ref{fig:delE_plot}(B) show the computed PMET rates as a function of the donor-bridge energy gap $\Delta E = U_{\mathrm{B}}-U_{\mathrm{D}}$ at a high cavity frequency ($\hbar \omega_{\mathrm{c}}$ = 200 meV) and at a low cavity frequency ($\hbar \omega_{\mathrm{c}}$ = 40 meV). We set  $g_{\mathrm{c}}$ and $\eta_{\mathrm{c}}$ to be equal, where $\hbar g_{\mathrm{c}} = \chi\mu_{\mathrm{DB}}$, $\hbar \eta_{\mathrm{c}} = \chi \mu_{\mathrm{BA}}$.

When the molecule is decoupled from the cavity, the electron transfer occurs via the $|\mathrm{D}\rangle \to |\mathrm{B}\rangle \to |\mathrm{A}\rangle$ superexchange pathway and one finds that the ET rate decreases with $\Delta E = U_{\mathrm{B}}-U_{\mathrm{D}}$. This is  understood by analyzing at the donor-acceptor coupling $V_{\mathrm{DA}}^{\mathrm{eff}}$, which is inversely proportional to the  donor-bridge energy gap (see Eq.~(6) of the main text). With a cavity, ET in mediated by two different coupling  pathways (as described in the main text)
\begin{itemize}
    \item {\bf Pathway 1:} ET from  photon dressed donor states $|\mathrm{D},n\rangle$ to photon dressed acceptor states $|\mathrm{A},m\rangle$ via bridge mediated channels $|\mathrm{B},l\rangle$.
    \item {\bf Pathway 2:} ET from direct transitions between $|\mathrm{D},n\rangle$ to $|\mathrm{A},m\rangle$. This direct transition is controlled by the cavity properties, such as the light-matter coupling strengths and the photon frequency.  
\end{itemize}
Figure~\ref{fig:delE_plot}(A) shows the computed PMET rate for a high photon frequency ($\hbar \omega_{\mathrm{c}}$ = 200 meV, $\hbar \omega_{\mathrm{c}}\gg k_{\mathrm{B}}T$). In this regime, only $|\mathrm{D},0\rangle$ has appreciable thermal population. Hence, the  predominant reactive channels are $|\mathrm{D},0\rangle \to \sum_l|\mathrm{B},l\rangle \to \{|\mathrm{A},0\rangle,|\mathrm{A},1\rangle\}$ via pathways (1), (bridge mediated), and $|\mathrm{D},0\rangle \to \{|\mathrm{A},0\rangle,|\mathrm{A},1\rangle\}$ via pathways (2), (direct transitions from donor to acceptor). The higher energy acceptor states $|\mathrm{A},2\rangle$, $|\mathrm{A},3\rangle$... are energetically disfavored due to their reduced Franck-Condon factors. At relatively low $\Delta E$ ($\sim$ 0.5-1 eV), these two pathway interfere destructively, so the overall rate is smaller compared to the cavity free scenario. Increasing $\Delta E$ causes the charge transfer to occur mostly via pathway (2), so the destructive interference is reduced and the reaction rate is enhanced. This rate  enhancement is proportional to the light-matter coupling strength, as indicated in the figure. The rate enhancement can also be understood from  Eq.~\ref{eq.total_DA_coup}. The first term in Eq.~\ref{eq.total_DA_coup} describes the direct transition from donor to acceptor, and this term grows with increasing light-matter coupling. For smaller $\Delta E$ values, the ET rate decreases due to the larger destructive interference between pathways (1) and (2). Strong light-matter coupling can overcome this destructive interference and boost the overall rate. For instance, Figure~\ref{fig:delE_plot}(A) shows that at $\Delta E$ = 0.5-1.0 eV, and $g_\mathrm{c} = \eta_{c} >$ 2 meV, the rates are largely enhanced in comparison with the case of $g_\mathrm{c} = \eta_{c} \le$ 2 meV. 
 
Figure~\ref{fig:delE_plot}(B) shows the PMET rate at low photon frequency ($\hbar \omega_{\mathrm{c}} = 40$ meV, $\hbar \omega_{\mathrm{c}}\ll k_{\mathrm{B}}T$). The fundamental difference is that the high lying donor dressed states (i.e., $|\mathrm{D},1\rangle$, $|\mathrm{D},2\rangle$, $|\mathrm{D},3\rangle$...) are initially populated or have appreciable thermal population. Thus, these high-lying channels can participate in the ET reaction. The significant difference is that, at low photon frequency, the destructive interference is larger compare to the case with high photon frequencies. For example, for comparable effective light-matter coupling strengths ($g_{\mathrm{c}}/\omega_{\mathrm{c}}$), with $g_\mathrm{c}$ = 4.0 meV ($g_{\mathrm{c}}/\omega_{\mathrm{c}}\simeq 0.02$) in Figure~\ref{fig:delE_plot}(A) and $g_{\mathrm{c}}$ = 1.0 meV ($g_{\mathrm{c}}/\omega_{\mathrm{c}}\simeq 0.025$) in Figure~\ref{fig:delE_plot}(B), the overall rate profile for Figure~\ref{fig:delE_plot}(B) is much smaller in magnitude compared to the data in Figure~\ref{fig:delE_plot}(A). This difference in the rate profiles arises from the multiple-donor dressed states that induce destructive interference for low photon frequency (Figure~\ref{fig:delE_plot}(B)) that is more pronounced over a wide range of $\Delta E$.

\bibliography{reference.bib}